\documentclass[1p,sort&compress,times]{elsarticle}
\usepackage{amsmath,amssymb,amscd,genyoungtabtikz,mathdots}
\usepackage{color}
\makeatletter
\def\ps@pprintTitle{%
 \let\@oddhead\@empty
 \let\@evenhead\@empty
 \def\@oddfoot{}%
 \let\@evenfoot\@oddfoot}
\makeatother
\newcommand{\al}{\alpha}
\newcommand{\be}{\beta}
\newcommand{\de}{\delta}

\newcommand{\vep}{\varepsilon}
\newcommand{\ga}{\gamma}
\newcommand{\ka}{\kappa}
\newcommand{\la}{\lambda}

\newcommand{\si}{\sigma}

\newcommand{\vp}{\varphi}
\newcommand{\ze}{\zeta}
\newcommand{\De}{\Delta}

\newcommand{\bk}{\mathbf{k}}

\newcommand{\bs}{\mathbf{s}}

\newcommand{\bde}{{\boldsymbol{\de}}}

\newcommand{\ta}{\tilde a}

\newcommand{\tmu}{\tilde\mu}

\newcommand{\NN}{{\mathbb N}}
\newcommand{\QQ}{{\mathbb Q}}
\newcommand{\RR}{{\mathbb R}}
\newcommand{\CC}{{\mathbb C}}
\newcommand{\ZZ}{{\mathbb Z}}
\newcommand{\cE}{{\mathcal E}}

\newcommand{\cH}{{\mathcal H}}

\newcommand{\cN}{{\mathcal N}}
\newcommand{\cP}{{\mathcal P}}

\newcommand{\id}{1\hspace{-.25em}{\rm I}}

\newcommand\ket[1]{|#1\rangle}
\newcommand\bra[1]{\langle#1|}

\newcommand{\ms}{\mspace{1mu}}
\renewcommand{\le}{\leqslant}
\renewcommand{\ge}{\geqslant}

\newcommand{\iu}{\mathrm{i}}
\newcommand{\e}{\mathrm{e}}
\newcommand{\Or}{\mathrm{O}}
\newcommand{\diff}{\mathrm{d}}

\newcommand{\gl}{\mathrm{gl}}
\newcommand{\su}{\mathrm{su}}
\newcommand{\sn}{\operatorname{sn}}
\newdefinition{remark}{Remark}
\newdefinition{example}{Example}
\begin{document}
\begin{frontmatter}
  \title{Yangian-invariant spin models and Fibonacci numbers}

  \author{Federico Finkel}
  \ead{ffinkel@ucm.es}
  
  \author{Artemio Gonz\'alez-L\'opez\corref{cor}}
  \ead{artemio@ucm.es}
  
  \address{Departamento de F\'\i sica Te\'orica II, Universidad Complutense de Madrid, 28040
    Madrid, Spain}
  \begin{abstract}
    We study a wide class of finite-dimensional $\su(m|n)$-supersymmetric models closely related
    to the representations of the Yangian~$Y(\gl(m|n))$ labeled by border strips. We
    quantitatively analyze the degree of degeneracy of these models arising from their Yangian
    invariance, measured by the average degeneracy of the spectrum. We compute in closed form the
    minimum average degeneracy of any such model, and show that in the non-supersymmetric case it
    can be expressed in terms of generalized Fibonacci numbers. Using several properties of these
    numbers, we show that (except in the simpler $\su(1|1)$ case) the minimum average degeneracy
    grows exponentially with the number of spins. We apply our results to several well-known spin
    chains of Haldane--Shastry type, quantitatively showing that their degree of degeneracy is
    much higher than expected for a generic Yangian-invariant spin model. Finally, we show that
    the set of distinct levels of a Yangian-invariant spin model is described by an effective
    model of quasi-particles. We study this effective model, discussing its connections to
    one-dimensional anyons and properties of generalized Fibonacci numbers.
\end{abstract}
\end{frontmatter}

\section{Introduction}\label{sec.intro}

In this paper we shall consider a general class of finite-dimensional quantum models which by
construction are invariant under the Yangian algebra $Y(\gl(m|n))$. The key to the Yangian
invariance of these models is their connection to certain finite-dimensional representations of
$Y(\gl(m|n))$ labeled by a type of skew Young diagrams, the so called border
strips~\cite{KKN97,NT95}. These representations play a fundamental role in the study of the
integrable two-dimensional conformal field theory related to the latter models, namely the
$\su(m|n)$ WZNW model at level 1~\cite{HHTBP92,Hi95npb,BS96}. The models we shall be interested
in, which we shall refer to as Yangian-invariant $\su(m|n)$ spin models, are characterized by the
fact that their Hilbert space is a direct sum of the irreducible representations of $Y(\gl(m|n))$
labeled by border strips with exactly $N$ boxes, where $N$ is the number of spins. The simplest
examples of these models are certain integrable spin chains with long range interactions invariant
under the Yangian for a finite number of spins, namely the $\su(m|n)$ supersymmetric versions of
the Haldane--Shastry~\cite{Ha88,Sh88,Ha93} and Polychronakos--Frahm~\cite{Po93,Fr93,HB00} chains;
see, e.g., \cite{BGHP93,HHTBP92,BS96,KKN97,BBH10}.

As is well known, the Haldane--Shastry (HS) spin chain is a circular (translationally invariant)
chain with equally spaced sites, the interactions between any two spins being inversely
proportional to the square of their chord distance. This model appears in connection with a
wide variety of topics in theoretical and mathematical physics, including one-dimensional
anyons~\cite{Ha91b,HHTBP92,Ha93,BS96}, conformal field theory~\cite{Ha91,BBS08,CS10,NCS11},
quantum chaos~\cite{FG05,BB06,BFGR08epl,BFGR09}, quantum information theory~\cite{GSFPA10}, and
quantum integrability~\cite{Ka92,HH93,BGHP93}. A distinctive feature of the HS chain is the fact
that it can be obtained from a dynamical model, namely the spin Sutherland (trigonometric)
model~\cite{Su71,Su71b,HH92}, in the strong coupling limit~\cite{Po93}. By applying the same
procedure to the spin Calogero (rational)~\cite{Ca71,MP93} and Inozemtsev (hyperbolic)~\cite{In96}
models, one obtains the Polychronakos--Frahm (PF) and Frahm--Inozemtsev (FI)~\cite{FI94} spin
chains, which share many properties with the original HS chain. In fact, although the Yangian
symmetry of the FI chain has not been rigorously established, we shall see in
Section~\ref{sec.Yinv} that it is isospectral to a Yangian-invariant spin model.

It is clear by its very definition that the class of Yangian-invariant spin models encompasses a
wide range of quantum systems. For instance, apart from the integrable spin chains mentioned
above, it includes the family of one-dimensional vertex models studied in Refs.~\cite{BBH10,BB12}.
A common feature shared by all Yangian-invariant spin models is the high degeneracy of their
spectrum, stemming from the Yangian symmetry. This statement, however, is far from precise, and
does not shed any light on whether a certain model in this class has additional degeneracy due to
its particular form. The main aim of this paper is precisely to perform a quantitative analysis of
the degeneracy inherent to all Yangian-invariant spin models. To this end, we shall use as a
concrete measure of the degree of degeneracy of a finite-dimensional quantum system its average
degeneracy, defined as the quotient of the dimension of its Hilbert space by the number
of distinct energy levels. We shall derive closed-form expressions for the average degeneracy of a
``generic'' Yangian-invariant $\su(m|n)$ spin model, both in the supersymmetric ($mn\ne0$) and
non-supersymmetric cases. In fact, these expressions provide a lower bound for the average
degeneracy of {\em any} Yangian-invariant spin model, and thus can be used as a
practical test for ruling out that a particular quantum system belongs to this class.

As it turns out, the behavior of this lower bound (which we have termed ``minimum average
degeneracy'') is rather different in the supersymmetric and non-supersymmetric cases. In the
former case, the minimum average degeneracy is given by a simple power law in terms of the
dimension $m+n$ of the one-particle Hilbert space (cf.~Eq.~\eqref{sumnbd}). Far more
interestingly, in the non-supersymmetric case the minimum average degeneracy is expressed in terms
of generalized Fibonacci numbers~\cite{Mi60}, which reduce to the standard ones for spin $1/2$
($m$ or $n$ equal to $2$). Using standard properties of the generalized Fibonacci numbers, we
derive the asymptotic behavior as the number of spins tends to infinity of the minimum average
degeneracy in the non-supersymmetric case. We find that the leading behavior is again a power law
involving the largest root (in module) of the characteristic polynomial of the generalized
Fibonacci numbers.

Another goal of this work is to ascertain to what degree the average degeneracy of the three spin
chains of Haldane--Shastry type (i.e., the HS, PF and FI chains) behaves as expected for a generic
Yangian-invariant spin model. Contrary to popular belief, we find out that this is actually not
the case. More precisely, we show that the number of distinct levels of these chains grows
polynomially with the number of spins, whereas for generic Yangian-invariant spin models this
growth is exponential. As explained in detail in Section~\ref{sec.avdeg}, the ultimate reason for
this different behavior is the fact that spin chains of HS type possess a polynomial dispersion
relation. As a matter of fact, our results apply as well to the whole family of vertex models
studied in Refs.~\cite{BBH10,BB12}, which also have this property. Moreover, when the coefficients
of a polynomial dispersion relation are rational numbers (as is the case for the HS and PF
chains, and for the FI chain when its parameter is rational) we show that the number of distinct
levels behaves as $N^{k+1}$, where $k$ is the total degree of the dispersion relation.

Another aspect of Yangian-invariant spin models that we shall analyze in this paper is their
connection with one-dimensional anyons. In fact, it is well known that spin chains of HS type
provide one of the simplest realizations of anyons in one dimension via Haldane's fractional
statistics~\cite{Ha91b}. For instance, as shown in the latter reference, the spectrum of the
$\su(2)$ HS chain is the same as that of a system of spin $1/2$ anyons with exclusion parameter
$g=1/2$. This result was essentially extended to the $\su(m)$ case in~\cite{BS96}, with $g=1/m$.
In this paper we focus instead on the set of {\em distinct} energy levels of an $\su(m)$
Yangian-invariant spin model, and show that it can be obtained from the spectrum of an effective
model of quasi-particles corresponding to the $1$'s in a Haldane motif~\cite{HHTBP92}. In
particular, in the $\su(2)$ case this effective model is simply a system of anyons with $g=2$. In
the general $\su(m)$ case we compute in closed form the statistical weights of these effective
models, and use them to derive in a novel way several combinatorial identities for generalized
Fibonacci numbers.

This paper is organized as follows. In Section~\ref{sec.Yinv} we review the description of the
representations of the Yangian $Y(\gl(m|n))$ labeled by border strips, and give a precise
definition of Yangian-invariant $\su(m|n)$ spin models. We also recall the definitions of the
$\su(m|n)$ spin chains of HS type, and discuss their spectra and dispersion relations.
Section~\ref{sec.adgFn} is devoted to the computation of the minimum average degeneracy of an
arbitrary Yangian-invariant spin model, in both the supersymmetric and non-supersymmetric cases.
We apply our results to the $\su(m|n)$ supersymmetric version of Inozemtsev's elliptic
chain~\cite{In90}, showing that in the $\su(2)$, $\su(3)$ and $\su(2|1)$ (or equivalently
$\su(1|2)$) cases it is {\em not} a Yangian invariant spin model for a wide range of values of
$N$. Using several properties of the generalized Fibonacci numbers, we also determine the
asymptotic behavior of the minimum average degeneracy of a Yangian-invariant spin model. In
Section~\ref{sec.avdeg-tim} we study the average degeneracy of Yangian-invariant spin models that
are also translationally invariant, like the Haldane--Shastry chain. As an example, we consider
the $\su(1|1)$ Inozemtsev chain, which is known to be (isospectral to) a Yangian- and
translationally-invariant spin model~\cite{FG14JSTAT}. As mentioned above, in
Section~\ref{sec.avdeg} we present a detailed analysis of the average degeneracy of spin chains of
HS type and, more generally, of Yangian-invariant spin models with a polynomial dispersion
relation. In Section~\ref{sec.anyons} we introduce the effective models describing the set of
distinct levels of a non-supersymmetric Yangian-invariant spin model, and discuss their
interpretation in terms of anyons and their connection with several identities satisfied by
generalized Fibonacci numbers. The paper ends with a brief section in which we summarize our main
results.

\section{Yangian-invariant spin models}\label{sec.Yinv}

In this section we shall provide a precise definition of the class of Yangian-invariant models on
which we shall focus in this paper. For the reader's convenience, we shall begin by recalling the
precise definition of the Yangian~$Y(\gl(m|n))$. Following the original paper~\cite{Na91} and
Ref.~\cite{BBHS07}, we first give an explicit matrix realization of the Lie
superalgebra~$\gl(m|n)$. To this end, let us introduce a $\ZZ_2$ grading~$p$ in the vector space
$\CC^{m+n}$ by setting $p(v)=0$ if $v\in\CC^m\times\{0\}$ and $p(v)=1$ if $v\in\{0\}\times\CC^n$. This
induces a natural grading in the space~$M_{m+n}(\CC)$ of $(m+n)\times(m+n)$ complex matrices given by
\[
\deg(E^{\al\be})=p(\al)+p(\be)\,,
\]
where~$p(\al)\equiv p(e_\al)$, $\{e_1,\dots,e_{m+n}\}$ being the canonical basis of $\CC^{m+n}$,
and $E^{\al\be}$ denotes the matrix whose only nonzero entry is a~$1$ in the $\al$-th row and
$\be$-th column. By definition, the Lie superalgebra~$\gl(m|n)$ is the vector
space~$M_{m+n}(\CC)$ with the
the graded commutator defined by the customary formula
\[
[A,B]_\pm=AB-(-1)^{\deg A+\deg B}BA
\]
on matrices with well-defined degree, and extended by linearity to the whole space. The
Yangian~$Y(\gl(m|n))$ is then defined by means of the solution of the Yang--Baxter equation given
by the $R$-matrix
\begin{equation}\label{Rmatrix}
R(u)=u-hP\,,
\end{equation}
where~$u$ is the spectral parameter, $h\in\CC$ and~$P$ is the graded permutation operator defined
by
\[
P(e_\al\otimes e_\be)=(-1)^{p(\al)p(\be)}e_\be\otimes e_\al\,.
\]
More precisely, $Y(\gl(m|n))$ is the associative (complex) superalgebra with generators
\[
T_{\al\be}^{(k)}\,,\qquad \al,\be=1\,,\dots\,,m+n\,,\quad k=1,2,\dots\,,
\]
defined as follows. Let
\[
T(u)=\sum_{\al,\be=1}^{m+n}T_{\al\be}(u)E^{\al\be}\in M_{m+n}\bigl(Y(\gl(m|n))\bigr)\,,
\]
where the matrix elements $T_{\al\be}(u)$ are defined by
\begin{equation}\label{Ygen}
T_{\al\be}(u)=\de_{\al\be}+\sum_{k=1}^\infty T_{\al\be}^{(k)}\,u^{-k}\,.
\end{equation}
We also introduce a $\ZZ_2$ grading in $Y(\gl(m|n))$ by setting
\[
\deg\bigl(T_{\al\be}^{(k)}\bigr)=p(\al)+p(\be)\,,
\]
so that the matrix~$T(u)$ and the identity matrix $\id=\sum_{\al=1}^{m+n}E^{\al\al}$ are both
even. Given two even $(m+n)\times(m+n)$ matrices~$A$, $B$ with matrix elements in~$Y(\gl(m|n))$, we
define their graded tensor product by
\[
(A\otimes B)_{ij,kl}=(-1)^{(p_i+p_k)p_j}A_{ik}B_{kl}\,.
\]
The defining relation for the generators~$T_{\al\be}^{(k)}$ are then given by the equation
\[
R(u-v)\vphantom T^1T(u)\,\vphantom T^2T(v)=\vphantom T^2T(v)\,\vphantom T^1T(u)R(u-v)\,,
\]
where the matrix~$R$ is defined by Eq.~\eqref{Rmatrix} and we have used the standard
notation~$\vphantom T^1T(u)=T(u)\otimes\id$, $\vphantom T^2T(u)=\id\otimes T(u)$. Using the above
definition of graded tensor product, it is easy to check that the latter equation is equivalent to
the following system for the currents $T_{\al\be}(u)$ of the Yangian~$Y(\gl(m|n))$:
\begin{align*}
&(u-v)\big[T_{\al\be}(u),T_{\ga\de}(v)\big]_\pm=(-1)^{p_\al p_\ga+p_\be p_\ga+p_\al p_\be}\,h\cdot
\Big(T_{\ga\be}(u)T_{\al\de}(v)-T_{\ga\be}(v)T_{\al\de}(u)\Big)\,;\\
&\al\,,\be\,,\ga\,,\de=1,\dots,m+n\,.
\end{align*}
This system obviously determines the defining relations of the generators~$T_{\al\be}^{(k)}$ of
the Yangian superalgebra~$Y(\gl(m|n))$ through Eq.~\eqref{Ygen}. In fact, it is well known that
this algebra is actually generated by its lowest two generators~$T_{\al\be}^{(0)}$ and
$T_{\al\be}^{(1)}$.

\begin{figure}[htp]
  \centering
  \includegraphics{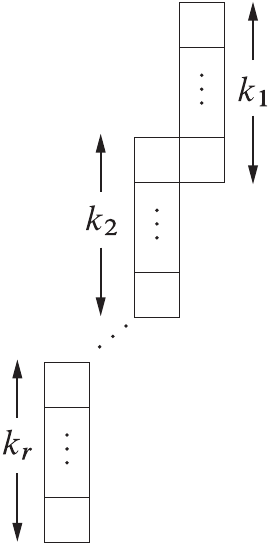}
  \caption{Border strip $(k_1,k_2,\dots,k_r)$.}
  \label{fig.bs}
\end{figure}

As mentioned in the Introduction, the models we shall deal with in this paper are associated to
finite-dimensional representations of the Yangian $Y(\gl(m|n))$ labeled by a particular class of
skew Young diagrams, namely border strips, that we shall describe next. We shall start with a few
preliminary definitions, following for the most part the notation in Ref.~\cite{KKN97}. A {\em
  skew Young diagram} $\la/\mu$ is the set of boxes obtained by subtracting a Young diagram $\mu$
from a larger Young diagram $\la\supset\mu$. Such a diagram $\la/\mu$ is {\em connected} if for
any pair of boxes $a,b\in\la/\mu$ there is a sequence of adjacent boxes starting with $a$ and
ending with $b$. A {\em border strip} is a connected skew Young diagram not containing any $2\times 2$
block of boxes. In other words, a border strip is a collection of boxes arranged in columns, such
that the top box in each column is adjacent to the bottom box in the column to its right. Thus a
border strip is determined by a vector $\bk\equiv(k_1,\dots,k_r)$, where $k_i$ is the number of
boxes in the $i$-th column, counting from right to left; cf.~Fig.~\ref{fig.bs}. For this reason,
from now on we shall usually identify a border strip with its corresponding vector of column
lengths $\bk$.
An $\su(m|n)$ \emph{semi-standard tableau} (SST) is a particular filling of the boxes in a border
strip $\bk$ with integers in the range $-n,-n+1,\dots,m-1$ according to the following two rules:
\begin{enumerate}
\item The integers in each column (read from top to bottom) and each row (read from left to right)
  form a nondecreasing sequence.
\item The nonnegative (respectively negative) integers in each row (resp.~column) form an
  increasing sequence.
\end{enumerate}
\begin{remark}
  The use of the term~$\su(m|n)$ in the previous definition is motivated by the fact that, as we
  shall explain in detail below, these tableaux appear in a natural way in the description of the
  spectrum of several well known spin chains possessing $\su(m|n)$ supersymmetry like, e.g., the
  Haldane--Shastry chain.
\end{remark}

\begin{figure}[hb]
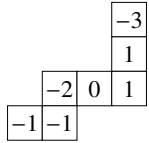

  \centering\small
  \gyoung(:::;{-3},:::;1,:;{-2};0;1,;{-1};{-1}) 
  \caption{$\su(2|3)$ SST with spin configuration $\bs=(-3,1,1,0,-2,-1,-1)$.}
  \label{fig.SST}
\end{figure}

We can obviously identify an $\su(m|n)$ SST with the sequence
\[
\bs\equiv(s_1,\dots,s_N)\,,\qquad s_i\in\{-n,-n+1,\dots,m-1\}\,,
\]
of its integers read from top to bottom and from right to left; we shall call such a sequence the
tableau's \emph{spin configuration}. For instance, the tableau in Fig.~\ref{fig.SST} is an
$\su(2|3)$ (or more generally $\su(m|n)$, with $m\ge2$ and $n\ge3$) SST with spin configuration
$\bs=(-3,1,1,0,-2,-1,-1)$, associated to the border strip $\bk=(3,1,2,1)$. We shall write
$\bs\in\bk$ to indicate that $\bs$ is the spin configuration of an $\su(m|n)$ SST whose underlying
border strip is $\bk$. We shall also say that two spin configurations $\bs$, $\bs'$ are
\emph{equivalent}, and use the notation $\bs\sim\bs'$, provided that $\bs,\bs'\in\bk$ for a
certain border strip $\bk$.

As is well-known, the set of all border strips is in one-to-one correspondence with an important
class of finite-dimensional irreducible representations of the Yangian $Y(\gl(m|n))$ whose carrier
spaces (the so called \emph{tame modules}) are irreducible under the action of a distinguished
maximal abelian subalgebra of $Y(\gl(m|n))$~\cite{NT95,KKN97}. Given a border strip $\bk$, the
dimension of the corresponding irreducible representation is the number of all possible $\su(m|n)$
SST associated with $\bk$. It is clear that any vector~$\bs\in\{-n,-n+1,\dots,m-1\}^N$ can be
regarded as the spin configuration of the $\su(m|n)$ tableau constructed by placing the first
component $s_1$ at the top of the first (rightmost) column, and then placing each component $s_i$
in the box below (respectively to the left) of the $(i-1)$-th box if $s_i>s_{i-1}$, or
$s_i=s_{i-1}\ge0$ (resp.~$s_i<s_{i-1}$, or $s_i=s_{i-1}<0$). It immediately follows that the
number of all $\su(m|n)$ SST with $N$ boxes coincides with the number $(m+n)^N$ of all spin
configurations $\bs\in\{-n,-n+1,\dots,m-1\}^N$. This implies that
\[
\dim\bigoplus_{\bk\in\cP_N}V_\bk(m|n)=(m+n)^N\,,
\]
where $V_\bk(m|n)$ denotes an irreducible $Y(\gl(m|n))$-module associated with the border strip
$\bk$ and $\cP_N$ is the set of all partitions of the integer $N$ (with order taken into account).
Thus the direct sum $\bigoplus_{\bk\in\cP_N}V_\bk(m|n)$ has the same dimension as the Hilbert
space of a spin chain with $N$ sites and two species of particles, one with $m$ and the other with
$n$ internal degrees of freedom. In fact, there is an important class of Yangian-invariant models
whose Hilbert space decomposes as the direct sum $\bigoplus_{\bk\in\cP_N}V_\bk(m|n)$, namely the
$\su(m|n)$-supersymmetric Haldane--Shastry and Polychronakos--Frahm
chains~\cite{Po94,BGHP93,Hi95npb,KKN97,HB00,BB06} that we shall now briefly describe.

To begin with, the Hilbert space of these chains is spanned by the canonical spin basis
\[
\ket{s_1}\otimes\cdots\otimes\ket{s_N}\equiv\ket{s_1,\dots,s_N}\equiv\ket\bs\,,\qquad
s_i\in\{-n,-n+1,\dots,m-1\}\,,
\]
where the $m$ bosonic (resp.~$n$ fermionic) internal degrees of freedom of the $i$-th spin
correspond to nonnegative (resp. negative) values of $s_i$. The action of the
$\su(m|n)$-supersymmetric spin permutation operators $S_{ij}^{(m|n)}$ (with $i<j$) on the
canonical spin basis is then defined by
\[
S_{ij}^{(m|n)}\,\ket{s_1,\dots,s_i,\dots,s_j,\dots,s_N}
=\si_{ij}(\bs)\,\ket{s_1,\dots,s_j,\dots,s_i,\dots,s_N}\,,
\]
with $\si_{ij}(\bs)=-1$ when $s_i$ and $s_j$ are both fermionic, or when $s_i$ and $s_j$ are of
opposite type and the number of fermionic spins occupying the sites $i+1,\dots,j-1$ is odd. In
terms of these operators, the Hamiltonians of the $\su(m|n)$ HS and PF chains can be taken
as~\cite{BUW99,BB06}
\begin{equation}
  \label{HS-PF}
  H = \sum_{1\le i<j\le N}J_{ij}\Big(1-S_{ij}^{(m|n)}\Big)\,,
\end{equation}
where
\begin{equation}\label{Jij}
J_{ij}=
\begin{cases}
  \frac12\sin^{-2}\bigl(\pi(i-j)/N\bigr)\,,\quad&\text{for the HS chain,}\\[3pt]
  (\xi_i-\xi_j)^{-2}\,,&\text{for the PF chain,}
\end{cases}
\end{equation}
and $\xi_1<\cdots<\xi_N$ are the zeros of the Hermite polynomial of degree $N$.

For the HS and PF chains, there is a simple explicit formula for the energy associated with the
irreducible representation labeled by a border strip $\bk\in\cP_N$ that we shall now recall. To
this end, it is convenient to introduce an alternative notation for the border strips in terms of
the so-called motifs~\cite{Ha91,HHTBP92}. More precisely, to a given a border strip
$\bk\in\cP_N$ we shall associate the vector $\bde\in\{0,1\}^{N-1}$ with components
\[
\de_i=
\begin{cases}
  1\,,\quad &i=k_1,\,k_1+k_2,\,\dots\,,\,k_1+\dots +k_{r-1},\\[3pt]
  0\,,&\text{otherwise.}
\end{cases}
\]
We shall call this vector $\bde$ the \emph{motif}\footnote{In the original definition of Haldane,
  the motif is the sequence of the components of $\bde$ with a zero added at both ends.}
representing the border strip~$\bk$, and shall refer to the positions
$K_i\equiv k_1+k_2\cdots+k_i$ of its nonzero components as its \emph{rapidities}. For instance,
for the border strip $\bk=(3,1,2,1)\in\cP_7$ in Fig.~\ref{fig.SST} the rapidities are $3,4,6$, and
thus the corresponding motif is $(0,0,1,1,0,1)$. For a given $\su(m|n)$ SST with spin
configuration $\bs$, we define $\bde(\bs)$ as the motif $\bde$ of the border strip determined by
$\bs$. It is immediate to convince oneself that
\[
\de_i(\bs)=\begin{cases} 1,\quad& \text{if}\ s_{i+1}<s_{i}\,\text{ or }s_i=s_{i+1}<0\,,\\[3pt]
  0\,,&\text{otherwise.}
\end{cases}
\]
With these definitions, the spectrum of the $\su(m|n)$ HS and PF chains with $N$ sites (with the
correct degeneracy for each level) can be shown to be the set of numbers~\cite{HB00,BBH10}
\begin{equation}
  \label{HSPFspec}
  \cE_N(\bs)=\sum_{j=1}^{N-1}\vep_N(j)\,\de_j(\bs)\,,
\end{equation}
with {\em dispersion relation}
\begin{equation}
  \label{cEN}
  \vep_N(j)=
  \begin{cases}
    j(N-j)\,,\quad&\text{for the HS chain,}\\
    j\,,&\text{for the PF chain.}
  \end{cases}
\end{equation}
In fact, a similar result holds for the (purely bosonic or fermionic) $\mathrm{su}(m)$
Frahm--Inozemtsev (FI) chain~\cite{FI94}, whose Hamiltonian is of the form~\eqref{HS-PF} with
couplings
\[
J_{ij}=\frac12\,\sinh^{-2}(\ze_i-\ze_j)\,.
\]
Here $\e^{2\ze_k}$ denotes the $k$-th root of the generalized Laguerre polynomial $L^{\al-1}_N$, and
$\al>0$ is a free parameter. In this case the spectrum is also given by Eq.~\eqref{HSPFspec}, with dispersion
relation~\cite{BFGR10}
\begin{equation}\label{FI}
\vep_N(j)=j(\al+j-1)\,.
\end{equation}
It can be shown that this formula for the spectrum holds as well in the $\su(m|n)$ case. In view
of these facts, it is natural to conjecture that the $\su(m|n)$ FI chain is also invariant under
the Yangian $Y(\gl(m|n))$, although to the best of our knowledge this result has not been
rigorously proved.

We shall study in this paper the degeneracy of the spectrum of the class of
$Y(\gl(m|n))$-invariant quantum systems whose Hilbert space $\cH_N$ decomposes as the direct sum
\begin{equation}
  \label{cHN}
  \cH_N=\bigoplus_{\bk\in\cP_N}V_{\bk}(m|n)
\end{equation}
of the irreducible representations of $Y(\gl(m|n))$ labeled by border strips with exactly $N$
boxes. We shall call any such system a {\em Yangian-invariant $\su(m|n)$ spin model} with $N$
sites. As we have just seen, this class includes the $\su(m|n)$-supersymmetric HS and PF
long-range spin chains. The Yangian invariance of these spin models and the
decomposition~\eqref{cHN} imply that their spectrum is of the form
\[
\cE_N(\bs)\,,\qquad \bs\in\{-n,-n+1,\dots,m-1\}^N\,,
\]
where $\cE_N$ is a real-valued function satisfying
\[
\bs\sim\bs'\implies \cE_N(\bs)=\cE_N(\bs')\,.
\]
Indeed, two spin configurations are equivalent if and only if they determine the same border strip
$\bk$, which labels a unique irreducible representation $V_\bk(m|n)$ of the Yangian. Since
obviously
\[
\bs\sim\bs'\iff\bde(\bs)=\bde(\bs')\,,
\]
we must have
\[
\cE_N(\bs)=E_N\bigl(\bde(\bs)\bigr)\,.
\]
Thus the spectrum of a Yangian-invariant spin model can be equivalently described by the function
$E_N:\{0,1\}^{N-1}\to\RR$ as the set of numbers
\begin{equation}\label{tEN}
E_N\bigl(\bde(\bs)\bigr)\,,\qquad \bs\in\{-n,-n+1,\dots,m-1\}^N\,.
\end{equation}
We shall call $E_N$ the \emph{energy function} of the system. It follows from Eq.~\eqref{tEN}
that the Hamiltonian of a Yangian-invariant spin model can be expressed as
\[
H=\sum_{\bk\in\cP_N}E_N(\bde)\sum_{\bs\in\bk}\ket{\bs}\bra{\bs}\,,
\]
where $\big\{\,\ket{\bs}\,|\,\bs\in\bk\,\big\}$ is any orthonormal basis of the subspace
$V_{\bk}(m|n)$ and $\bde$ denotes the motif corresponding to the border strip $\bk$. For instance,
for the $\su(m|n)$ HS and PF chains the energy function is the linear functional
\begin{equation}\label{linfunc}
E_N(\bde)=\sum_{j=1}^{N-1}\vep_N(j)\de_j\,.
\end{equation}
with~$\vep_N$ given by Eq.~\eqref{cEN}. In fact, more general Yangian-invariant spin models having
a linear energy function~\eqref{linfunc} with polynomial dispersion relation $\vep_N(j)$ have
recently been studied in Refs.~\cite{BBH10,BB12}.

We shall take Eq.~\eqref{tEN} as the basis of our analysis of the spectrum of a
Yangian-invariant spin model. For this reason, our results will also apply to models like the FI
chain, whose Yangian invariance has not been established but which is known to possess a spectrum
of the form~\eqref{tEN} (with $E_N$ given by~\eqref{FI}-\eqref{linfunc}).

\section{Average degeneracy and generalized Fibonacci numbers}\label{sec.adgFn}

\subsection{Minimum average degeneracy}

The main goal of this section is to compute a lower bound for the \emph{average degeneracy}
of the spectrum~\eqref{tEN} of a Yangian-invariant spin model with $N$ sites, defined by
\begin{equation}
  \label{avdeg}
  d_N = \frac{(m+n)^N}{\ell_N}\,,
\end{equation}
where $\ell_N$ is the number of \emph{distinct} energies. This is of course equivalent to finding
an upper bound on $\ell_N$, a problem which we shall now address. The key observation in this
respect is to note that, by Eq.~\eqref{tEN}, $\ell_N$ is obviously equal to the number of distinct
values taken by the energy function $E_N(\bde)$, where $\bde$ ranges over the set
\begin{equation}
  \label{DeNmn}
  \De_N(m|n)=\big\{\bde(\bs)\mid \bs\in\{-n,-n+1,\dots,m-1\}^N\big\}\,.
\end{equation}
of all valid $\su(m|n)$ motifs with $N-1$ components. Thus
\begin{equation}\label{nuNbd}
\ell_N\le\nu_N(m|n)\,,
\end{equation}
where $\nu_N(m|n)$ denotes the cardinal of the set $\De_N(m|n)$, and therefore
\begin{equation}
  \label{dNbound}
  d_N\ge d_{N,\mathrm{min}}\,,
\end{equation}
where the {\em minimum average degeneracy} $d_{N,\mathrm{min}}$ is given by
\begin{equation}\label{dNmin}
 d_{N,\mathrm{min}}=\frac{(m+n)^N}{\nu_N(m|n)}\,.
\end{equation}
In fact, the minimum average degeneracy is strictly less than the average degeneracy if and only
if the energy function $E_N$ is not injective; we shall say in this case that the
Yangian-invariant spin model exhibits {\em accidental degeneracy}. We shall also say that a
Yangian-invariant spin model is {\em generic} if it has no accidental degeneracy.

As we have just seen, in order to evaluate $d_{N,\mathrm{min}}$ in closed form we only need to
compute the cardinal of the set~\eqref{DeNmn}. To this end, we shall next recall a simple
characterization of all allowed $\su(m|n)$ motifs which shall be of fundamental importance in what
follows. Consider, to begin with, the genuinely supersymmetric case $mn\ne0$. It is easy to
realize that in this case any sequence $\bde\in\{0,1\}^{N-1}$ is a valid motif, so that
\begin{equation}\label{nuNsuper}
\nu_N(m|n)=2^{N-1}\,,\qquad mn\ne0\,.
\end{equation}
Indeed, if $\bde\in\{0,1\}^{N-1}$ let $K_1,\dots,K_{r-1}$ be the positions of the nonzero
components of $\bde$, and define $k_i=K_i-K_{i-1}$ (with $K_0\equiv0$ and $K_r\equiv N$). Then the
vector
\[
\bs=(\underbrace{0,\dots,0}_{k_1},\underbrace{-1,0,\dots,0}_{k_2},\dots,
\underbrace{-1,0,\dots,0}_{k_r})
\]
is the spin configuration of an $\su(m|n)$ SST whose corresponding motif is $\bde$. Thus in this
case we just have
\begin{equation}\label{sumnbd}
  d_{N,\mathrm{min}}=2\,\bigg(\frac{m+n}2\bigg)^N\,,\qquad mn\ne0\,.
\end{equation}

Consider next the purely bosonic $\su(m|0)$ case. It is clear that an $\su(m|0)$ motif cannot
contain a sequence of $m$ or more consecutive 1's, since the spin configuration of any tableau
associated to such a motif would contain a sequence of $m+1$ or more distinct integers in the
range $0,1,\dots,m-1$. Conversely, suppose that the vector $\bde\in\{0,1\}^{N-1}$ contains no
sequence of $m$ or more consecutive 1's. Such a $\bde$ can be written as a succession of sequences
of consecutive 0's and 1's of the form
\[
\bde=(\underbrace{0,\dots,0}_{l_1},\underbrace{1,\dots,1}_{n_1},0,\dots)
\]
with $n_i\le m-1$. It is immediate to check that the spin configuration
\[
\bs=(\underbrace{m-1,\dots,m-1}_{l_1+1},m-2,\dots,m-n_1,m-n_1-1,m-1,\dots)\,,
\]
where each entry is in the range $0,1,\dots,m-1$ on account of the condition $n_i\le m-1$,
determines an $\su(m)$ SST whose corresponding motif is $\bde$. This shows that $\De_N(m|0)$
\emph{is the set of all vectors $\bde\in\{0,1\}^{N-1}$ containing no sequences of $m$ or more
  consecutive 1's.} In exactly the same way it is proved that in the purely fermionic case
$\De_N(0|n)$ is the set of all vectors $\bde\in\{0,1\}^{N-1}$ containing no sequences of $n$ or
more consecutive 0's. It immediately follows from these two facts that the sets $\De_N(m|0)$ and
$\De_N(0|m)$ are mapped bijectively into each other by the duality transformation
\begin{equation}\label{dual}
  \bde=(\de_1,\dots,\de_{N-1})\in\De_N(m|0)\mapsto \bde'=(1-\de_1,\dots,1-\de_{N-1})\in\De_N(0|m)\,.
\end{equation}
Hence
\begin{equation}\label{BF}
  \nu_N(m|0)=\nu_N(0|m),
\end{equation}
so that from now on we shall restrict ourselves without loss of generality to the purely bosonic
case.
\begin{remark}
  The duality transformation~\eqref{dual} actually arises from a mapping defined on spin
  configurations~\cite{BBHS07}, valid also in the general supersymmetric case. More precisely,
  given an $\su(m|n)$ spin configuration $\bs$ define the vector $\bs'\equiv(s_1',\dots,s_N')$ by
  \[
  s_i'=-s_i-1\,,\qquad i=1,\dots,N\,.
  \]
  It is immediate to check that $\bs'$ is a valid $\su(n|m)$ spin configuration, and that
  \begin{equation}\label{dedep}
    \bde(\bs')=\bde(\bs)'\,,
  \end{equation}
  where $\bde'$ is defined by Eq.~\eqref{dual}. Consider two $\su(m|n)$ and $\su(n|m)$ spin models
  with the same energy function~$E_N$. It is straightforward to show that the spectra of
  both models are mapped in a one-to-one way by the duality transformation
  \begin{equation}\label{Edual}
  E_N(\bde)\mapsto E_N(\bde')\,.
\end{equation}
Indeed, it is obvious from Eq.~\eqref{tEN} for the spectrum that the energies of both models are
in a one-to-one correspondence under the transformation~\eqref{Edual}. Their respective
degeneracies also coincide, since for a given $\su(m|n)$ motif $\bde$ we have
  \[
  \big\{\bs\in\{-n,-n+1,,\dots,m-1\}\mid\bde(\bs)
  =\bde\big\}=\big\{\bs'\in\{-m,-m+1,\dots,n-1\}\mid\bde(\bs')=\bde'\big\}\,,
  \]
  on account of Eq.~\eqref{dedep}. In particular, for the linear energy function~\eqref{linfunc}
  the relation between the corresponding spectra of two $\su(m|n)$ and $\su(n|m)$ models is simply
  given by
 \begin{equation}\label{Efb}
  E_{N}^{(n|m)}(\bde')=E_0-E_{N}^{(m|n)}(\bde)\,,
 \end{equation}
  where
\begin{equation}\label{Ezero}
  E_0=\sum_{j=1}^{N-1}\vep_N(j)
\end{equation}
  is the maximum energy of the model(s) with a nonvanishing number of fermionic degrees of
  freedom.
\end{remark}

In view of the above discussion, we need only compute the cardinal $\nu_N(m)\equiv\nu_N(m|0)$ of
the set of $\su(m)\equiv\su(m|0)$ (purely bosonic) motifs. The key observation in this respect is
that an $\su(m)$ motif $\bde$ of length $N-1\ge m$ must necessarily be of the form
\[
(\bde^{(k)},0,\underbrace{1,\dots,1}_{k-1})\,,\qquad k=1,\dots,m,
\]
where $\bde^{(k)}$ is a valid $\su(m)$ motif of length $N-k-1$. This observation immediately leads
to the recursion relation
\begin{subequations}\label{nuNm}
  \begin{equation}\label{recnu}
    \nu_N(m)=\sum_{k=1}^m\nu_{N-k}(m)\,,\qquad N\ge m+1\,.
  \end{equation}
  On the other hand, for $\su(m)$ motifs of length $N-1<m$ the restriction on the number of
  consecutive 1's is vacuous, so that
  \[
  \De_N(m|0)=\{0,1\}^{N-1}\,,\qquad N=1,\dots,m\,,
  \]
  and therefore
  \begin{equation}
    \label{incond}
    \nu_N(m)=2^{N-1}\,,\qquad N=1,\dots,m\,.
  \end{equation}
\end{subequations}
The recursion relation~\eqref{recnu}, together with the initial conditions~\eqref{incond},
uniquely determines $\nu_N(m)$ for arbitrary $N$. We shall next show that $\nu_N(m)$ can in fact
be expressed in terms of the so-called $m$-\emph{generalized Fibonacci} (in short,
\emph{$m$-nacci}) \emph{numbers}~$F^{(m)}_n$, defined~\cite{Mi60} as the unique solution of the
recursion relation
\begin{subequations}\label{FNm}
  \begin{equation}
    F^{(m)}_n=\sum_{k=1}^mF^{(m)}_{n-k}\,,\qquad n\ge m\,,
    \label{Fmrecrel}
  \end{equation}
  with the initial conditions
  \begin{equation}
    F^{(m)}_0=\dots=F^{(m)}_{m-2}=0,\quad F^{(m)}_{m-1}=1\,.
    \label{incondF}
  \end{equation}
\end{subequations}
Note, in particular, that for $m=2$ Eqs.~\eqref{FNm} yield the standard Fibonacci numbers.
Comparing Eqs.~\eqref{nuNm} with~\eqref{FNm} it is straightforward to prove that
\begin{equation}
  \label{nuF}
  \nu_N(m)=F^{(m)}_{N+m-1}\,.
\end{equation}
Indeed, it is obvious that the shifted $m$-nacci sequence $F_{N+m-1}^{(m)}$ satisfies the
recursion relation~\eqref{recnu}. Hence it suffices to show that~\eqref{nuF} holds for
$N=1,\dots,m$. This is obviously true for $N=1$. Assuming that~\eqref{nuF} also holds for
$N=1,\dots,k\le m-1$, by Eqs.~\eqref{FNm} we have
\[
F^{(m)}_{m+k}=F_k^{(m)}+\dots+F_{m-1}^{(m)}+\sum_{j=0}^{k-1}F^{(m)}_{m+j}=1+\sum_{j=0}^{k-1}\nu_{j+1}(m)
=1+\sum_{j=0}^{k-1}2^j=2^k=\nu_{k+1}(m)\,,
\]
so that~\eqref{nuF} also holds for $N=k+1$. By induction, the latter equation is true for
$N=1,\dots,m$, thus establishing our claim. From Eq.~\eqref{dNmin} with $n=0$ we conclude that
\emph{the minimum average degeneracy of a Yangian-invariant bosonic $\su(m)$ spin model is given by}
\begin{equation}\label{MAD}
d_{N,\text{min}}\equiv\frac{m^N}{F^{(m)}_{N+m-1}}\,.
\end{equation}
By Eq.~\eqref{BF}, the same is true for a purely fermionic $\su(m)$ spin model.

The inequality~\eqref{dNbound} provides a very simple necessary condition for a quantum system to
be a Yangian-invariant spin model, as we shall see in the following example.

\begin{example} Consider the $\su(m|n)$ supersymmetric version of the Inozemtsev
  elliptic chain~\cite{In90}, with Hamiltonian
\begin{equation}\label{H}
H^{(m|n)}=\frac12\sum_{1\le i<j\le N}\frac{1-S_{ij}^{(m|n)}}{\sn^2\bigl(2(i-j)\frac KN\bigr)}\,.
\end{equation}
In the latter equation $\sn$ denotes Jacobi's elliptic sine with real
period $4K$, where $K\equiv K(k)$ is the complete elliptic integral of the first kind
\begin{equation}\label{K}
K(k)=\int_0^{\pi/2}\frac{\diff x}{\sqrt{1-k^2\sin^2x}}
\end{equation}
and $k\in[0,1)$ is the elliptic modulus. As is well-known, for $k=0$ the latter model reduces to
the $\su(m|n)$ HS chain~\eqref{HS-PF}-\eqref{Jij} which, as we have seen in the previous section,
is a Yangian-invariant spin model. On the other hand, for $k\to1$ the Hamiltonian~\eqref{H} is
related to the $\su(m|n)$ Heisenberg chain, which does not possess Yangian invariance for a
finite-number of sites. For $k\in(0,1)$, it is generally believed that the Hamiltonian~\eqref{H}
is not invariant under $Y(\gl(m|n))$ for any finite value of $N$. We have numerically computed the
average degeneracy of the spectrum of the model~\eqref{H} with $\su(2)$, $\su(3)$ and $\su(2|1)$
spin%
\footnote{As shown in Ref.~\cite{BBHS07}, the operators $S_{ij}^{(m|n)}$ and $-S_{ij}^{(n|m)}$ are
  unitarily equivalent, so that the spectra of $H^{(m|n)}$ and $-H^{(n|m)}$ differ by an additive
  constant. In particular, the average degeneracies of the $\su(m|n)$ and $\su(n|m)$
  models~\eqref{H} coincide.} (for $6\le N\le 18$ in the first case and $6\le N\le 12$ in the
remaining ones). It turns our that for $k\in(0,1)$ the average degeneracy is practically
independent of $k$, so that we shall restrict ourselves to the case $k^2=1/2$. As can be seen from
Fig.~\ref{fig.avdegs}, for $N\ge 8$ the average degeneracy of the model~\eqref{H} in the cases
studied is clearly lower than the minimum average degeneracy of a Yangian-invariant spin model
(cf. Eqs.~\eqref{sumnbd} or \eqref{MAD}). This shows that in these cases the chain~\eqref{H} is
{\em not} a Yangian-invariant spin model (see also Ref.~\cite{FG14} for the $\su(2)$ case).
\begin{figure}[h]
    \centering
    \includegraphics[width=.475\columnwidth]{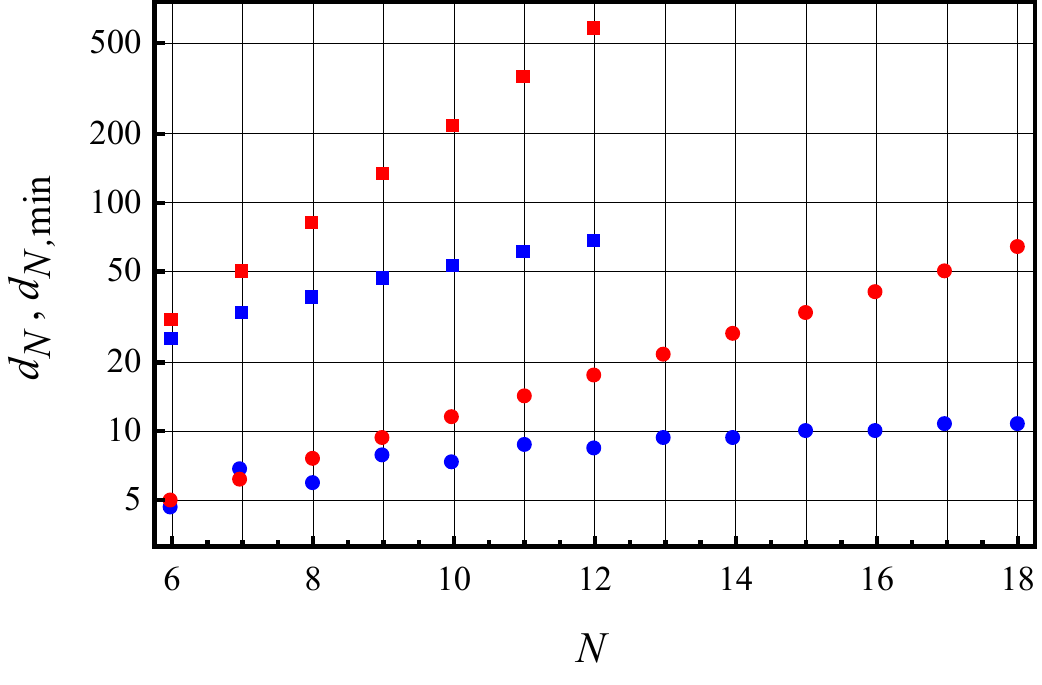}\hfill\includegraphics[width=.475\columnwidth]{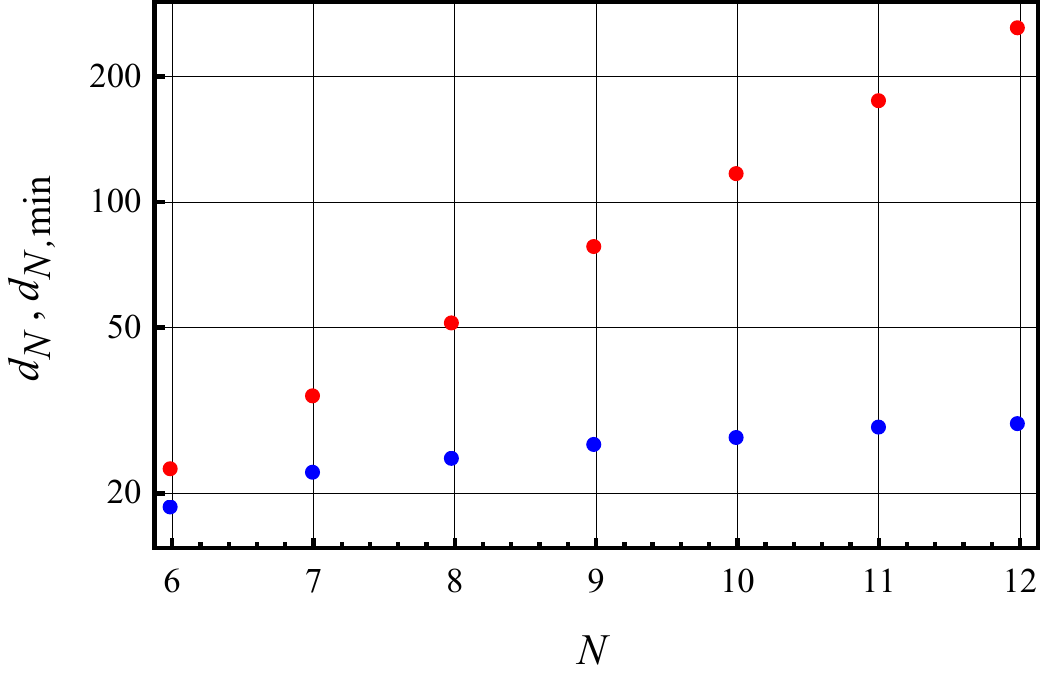}
    \caption{Left: logarithmic plot of the average degeneracy of the $\su(m)$ elliptic chain with
      $k^2=1/2$ and $m=2$ (blue circles) and $m=3$ (blue squares), compared to the minimum average
      degeneracy of an $\su(m)$ Yangian-invariant spin model with $m=2$ (red circles) and $m=3$
      (red squares). Right: analogous plot for the $\su(2|1)$ elliptic chain with $k^2=1/2$ (blue
      circles) and the $\su(2|1)$ minimum average degeneracy (red circles).}
    \label{fig.avdegs}
  \end{figure}
\end{example}

\subsection{Asymptotic behavior of the minimum average degeneracy}\label{abmad}

We shall next investigate the asymptotic behavior as $N$ tends to infinity of the minimum average
degeneracy~\eqref{MAD} of an $\su(m)$ Yangian-invariant spin model. To this end, note
that~\eqref{Fmrecrel} is a linear recursion relation with constant coefficients and
\emph{characteristic polynomial}
\begin{equation}
  \label{pm}
  p_m(\la)=\la^m-\sum_{k=1}^m\la^{m-k}\,.
\end{equation}
It is shown in Ref.~\cite{Mi60} that all the roots $\la_i^{(m)}$ ($i=1,\dots,m$) of $p_m(t)$ are
simple, and that moreover they can be labeled in such a way that
\begin{equation}
  \label{roots}
  \left|\la_i^{(m)}\right|<1\,,\quad i=1,\dots,m-1;\qquad 1<\la_m^{(m)}\equiv\la_m<2\,.
\end{equation}
From the simple character of the roots of the characteristic polynomial it follows that the
general solution of the recursion relation~\eqref{Fmrecrel} is of the form
\begin{equation}\label{gensol}
\sum_{i=1}^mc_i^{(m)}\left(\la_i^{(m)}\right)^n\,,\qquad n=0,1,\dots\,,
\end{equation}
where the coefficients $c_i^{(m)}$ are complex constants (independent of $N$). In order to obtain
the $m$-nacci numbers, the coefficients $c_i^{(m)}$ must be chosen in such a way
that~\eqref{gensol} satisfies the initial conditions~\eqref{incondF}, i.e.,
\[
\sum_{i=1}^mc_i^{(m)}\left(\la_i^{(m)}\right)^n=\de_{n,m-1}\,,\qquad n=0,\dots,m-1\,.
\]
Note that the coefficient matrix of this linear system is the Vandermonde matrix determined by the
$m$ distinct roots of the characteristic polynomial $p_m(\la)$, and therefore the system has
always a unique solution. In fact, it is straightforward to show that
\[
c_i^{(m)}=\prod_{j;j\ne i}\left(\la_i^{(m)}-\la_j^{(m)}\right)^{-1}\,,
\]
and therefore
\begin{equation}
  \label{Fmn}
  F^{(m)}_n=\sum_{i=1}^m\left(\la_i^{(m)}\right)^n
  \prod_{j;j\ne i}\left(\la_i^{(m)}-\la_j^{(m)}\right)^{-1}\,.
\end{equation}
From this equation and Eq.~\eqref{roots} it follows that as $n\to\infty$ we have
\begin{equation}
  \label{Fmnasymp}
  F_n^{(m)}=c_m\la_m^n+\Or(\e^{-\ka_mn})\,,
\end{equation}
with
\begin{equation}
  c_m\equiv c_m^{(m)}=\prod_{i=1}^{m-1}\left(\la_m-\la_i^{(m)}\right)^{-1}
  \label{cm}
\end{equation}
and
\[
\ka_m=-\min_{1\le i\le m-1}\log\left|\la_i^{(m)}\right|>0\,.
\]
Thus as $N\to\infty$ the minimum average degeneracy behaves as
\begin{equation}
  \label{dminas}
  d_{N,\text{min}}=\ga_m\bigg(\frac m{\la_m}\bigg)^N\Big(1+\Or\left(\e^{-(\ka_m+\log\la_m)N}\right)\Big)\,,
\end{equation}
where the coefficient $\ga_m$ (independent of $N$) is given by
\[
\ga_m=\frac1{c_m\la_m^{m-1}}=\prod_{i=1}^{m-1}\left(1-\frac{\la_i^{(m)}}{\la_m}\right)\,.
\]
Since $\la_m<2$ (cf.~Eq.~\eqref{roots}), it follows from Eq.~\eqref{dminas} that
$d_{N,\text{min}}$ grows exponentially as $N\to\infty$\,.

The coefficient $c_m$, and hence $\ga_m$, can be expressed in terms of the root $\la_m$. Indeed,
from Eq.~\eqref{cm} we have
\[
\frac1{c_m}=\prod_{i=1}^{m-1}\left(\la_m-\la_i^{(m)}\right)
=\lim_{\la\to\la_m}\frac{p_m(\la)}{\la-\la_m}
=p_m'(\la_m)\,.
\]
In particular, from the previous expression it is obvious that $c_m$ is real and positive, as it
should be on account of Eq.~\eqref{Fmnasymp}. Another useful expression for the coefficient $c_m$
is obtained by noting that, by Eq.~\eqref{pm},
\[
q_m(\la)\equiv(\la-1)p_m(\la)=\la^{m+1}-2\la^m+1
=(\la-1)(\la-\la_m)\prod_{i=1}^{m-1}\left(\la-\la_i^{(m)}\right)\,,
\]
and therefore
\begin{equation}\label{cmsimp}
\frac1{c_m}=\frac1{\la_m-1}\lim_{\la\to\la_m}\frac{q_m(\la)}{\la-\la_m}=\frac{q_m'(\la_m)}{\la_m-1}
=\frac{(m+1)\la_m^m-2m\la_m^{m-1}}{\la_m-1}\,.
\end{equation}
We thus arrive at the following remarkably simple expression for the coefficient $\ga_m$ in
Eq.~\eqref{dminas}:
\begin{equation}
  \ga_m=\frac{\la_m^{1-m}}{c_m}=\frac{(m+1)\la_m-2m}{\la_m-1}=m+1-\frac{m-1}{\la_m-1}\,.
  \label{gamsimp}
\end{equation}
In particular, from Eqs.~\eqref{roots} and~\eqref{gamsimp} we obtain the following upper and lower
bounds on $\la_m$:
\begin{equation}\label{lambds}
\frac{2m}{m+1}<\la_m<2\,,
\end{equation}
which imply that
\[
\lim_{m\to\infty}\la_m=2\,.
\]
In fact, it can be easily shown that the sequence $\{\la_m\}_{m=2}^\infty$ is monotonically
increasing. Indeed,
\[
p_{m+1}(\la_m)=p_{m+1}(\la_m)-p_{m}(\la_m)=\la_m^m(\la_m-2)<0\,,
\]
so that $\la_m<\la_{m+1}$ since $\la_{m+1}$ is the largest root of the monic polynomial $p_{m+1}$.

On the other hand, since $\la=1$ is not a root of $p_m$, the characteristic equation $p_m(\la)=0$
can be written as
  \begin{equation}\label{laeq}
  \la^m-\frac{\la^m-1}{\la-1}=0\,,
\end{equation}
or equivalently
\begin{equation}\label{lalamm}
\la=2-\frac1{\la^m}\,.
\end{equation}
Hence
\[
\lim_{m\to\infty}m(2-\la_m)=\lim_{m\to\infty}\frac{m}{\la_m^m}=0\,,
\]
and Eq.~\eqref{gamsimp} implies that
\[
\ga_m=\frac{\la_m-m(2-\la_m)}{\la_m-1}\underset{m\to\infty}{\longrightarrow}2\,.
\]
Thus, for $m\gg1$ we have
\[
d_{N,\text{min}}\simeq2\left(\frac m2\right)^N\,,\qquad m\gg1\,,
\]
where the RHS is the minimum average degeneracy of an $\su(p|q)$ model with $p+q=m$
(cf.~Eq.~\eqref{sumnbd}).

It can be numerically checked that $\la_m$ and (to a lesser extent) $\ga_m$ rapidly converge to
$2$ as $m$ tends to $\infty$ (see, e.g., Table~\ref{tab.laga} below). To confirm this observation
analytically, it suffices to note that from Eq.~\eqref{lambds} we have
\[
\frac1{\la_m^m}<\frac1{2^m}\,\left(1+\frac1m\right)^m<\frac{\e}{2^m}\,,
\]
and therefore, by Eq.~\eqref{lalamm},
\[
2-\frac{\e}{2^m}<\la_m<2\,.
\]
The upper bound on $\la_m$ implies that $\ga_m<2$, while from the lower bound we easily obtain
\[
\ga_m>m+1-\frac{m-1}{1-2^{-m}\e}=2-\frac{(m-1)\e}{2^m-\e}>2-\frac{m\e}{2^m}\qquad\text{for}\quad
m\ge4\,.
\]
\begin{remark}
  Equation~\eqref{cmsimp} for the coefficient $c_m$ can in fact be extended with exactly the same
  proof to the remaining coefficients $c_i^{(m)}$ ($i=1,\dots,m-1$), i.e.,
  \[
  c_i^{(m)}=\frac{\la_i^{(m)}-1}{\left(\la_i^{(m)}\right)^{m-1}\Big((m+1)\la_i^{(m)}-2m\Big)}\,.
  \]
  Inserting the latter equation into Eq.~\eqref{gensol} provides an alternative simple derivation
  of Binet's formula for the $m$-nacci numbers~\cite{SJ84,DD14}
  \[
  F_n^{(m)}=\sum_{i=1}^m\frac{\la_i^{(m)}-1}{(m+1)\la_i^{(m)}-2m}\,\left(\la_i^{(m)}\right)^{n-m+1}\,.
  \]
\end{remark}
\begin{example}
  For $m=2$ the roots of the characteristic polynomial $p_2(\la)$ are
  \[
  \la_1=\frac12\big(1-\sqrt 5\big)\,,\qquad \la_2=\frac12\big(1+\sqrt 5\big)\,,
  \]
  so that Eq.~\eqref{Fmn} reduces to the famous Binet formula for the Fibonacci numbers
  \[
  F_n^{(2)}\equiv F_n=\frac1{2^n\sqrt
    5}\,\Big[\left(1+\sqrt5\,\right)^n-\left(1-\sqrt5\,\right)^n\Big]\,.
  \]
  Thus the minimum average degeneracy~\eqref{MAD} of a Yangian-invariant $\su(2)$ spin model is
  given by
  \begin{equation}\label{dNmin2}
  d_{N,\text{min}}=\frac{2^{2N+1}\sqrt 5}{\left(1+\sqrt 5\,\right)^{N+1}-\left(1-\sqrt
      5\,\right)^{N+1}}\simeq\frac{\sqrt5}2\,\left(\!\sqrt 5-1\right)^{N+1}\,,
\end{equation}
in agreement with Eqs.~\eqref{dminas} and~\eqref{gamsimp}.

  In the case $m=3$, the characteristic polynomial has two complex conjugate roots of modulus less
  than one and a real root
  \[
  \la_3=\frac13\left(1+\sqrt[3]{19+3\sqrt{33}}+\sqrt[3]{19-3\sqrt{33}}\,\right)\equiv\frac13\,(1+r)
  \simeq 1.83929\,.
  \]
  Using Eq.~\eqref{gamsimp} we easily obtain
  \[
  \ga_3=\frac{2(2r-7)}{r-2}\simeq1.61702\,,
  \]
  so that for large $N$ the $\su(3)$ minimum average degeneracy behaves as
  \[
  d_{N,\text{min}}\simeq1.61702\cdot 1.63107^N\,,\qquad N\gg1\,.
  \]
  The exact formulas for $\la_4$ and $\ga_4$ are too unwieldy to display, and for $m\ge5$ the
  characteristic equation defining $\la_m$ cannot be solved in radicals. However, the numerical
  evaluation of $\la_m$ and $\ga_m$ is straightforward; see Table~\ref{tab.laga} for a list of
  their values in the range $m=2,\dots,10$.
 \begin{table}[h]
   \centering
   \begin{tabular}{|r|c|c|}
     \hline
     \vrule width 0pt height 10pt
     $m$&$\la_m$&$\ga_m$\\[2pt]
     \hline
     \vrule width 0pt height 12pt
    2& 1.61803& 1.38197\\
    3& 1.83929& 1.61702\\
    4& 1.92756& 1.76571\\
    5& 1.96595& 1.85899\\
    6& 1.98358& 1.91654\\
    7& 1.99196& 1.95139\\
    8& 1.99603& 1.97211\\
    9& 1.99803& 1.98420\\
     10& 1.99902& 1.99116\\[2pt]
     \hline
 \end{tabular}
\caption{Values of $\la_m$ and $\ga_m$ in the asymptotic formula~\eqref{dminas} for $m=2,\dots,10$.}
\label{tab.laga}
\end{table}
\end{example}

\section{Average degeneracy of translationally invariant models}\label{sec.avdeg-tim}

Several important examples of Yangian-invariant spin models possess the additional property of
being translationally invariant. This is notably the case for the $\su(m|n)$ Haldane--Shastry spin
chain~\eqref{HS-PF}-\eqref{Jij}, whose energy function is of the form~\eqref{linfunc} with
$\vep_N(j)=j(N-j)$. The dispersion relation of this chain obviously satisfies the
identity
\begin{equation}\label{tim}
  \vep_N(j)=\vep_N(N-j),
\end{equation}
characteristic of translationally invariant models (on a lattice with unit spacing between
consecutive sites) with a linear energy function. Indeed, for these models $\vep_N(j)$ is naturally
interpreted as the energy of a quasi-particle with momentum $2\pi j/N$. Since the momentum is
defined up to multiples of $2\pi$ due to the translation invariance, the identity~\eqref{tim}
simply expresses the fact that the energy does not depend on the sign of the momentum. When the
energy function $E_N(\bde)$ is linear (cf.~Eq.~\eqref{linfunc}) and its dispersion relation
satisfies Eq.~\eqref{tim}, the spectrum can be expressed as
\[
E_N(\bde)=\sum_{j=1}^{\lfloor(N-1)/2\rfloor}\vep_N(j)\big(\de_j+\de_{N-j}\big)+\big(1-\pi(N)\big)\,\vep_N(N/2)\,\de_{N/2}\,,
\qquad\bde\in\De_N(m|n)\,,
\]
where $\pi(N)$ is the parity of $N$ and $\lfloor\cdot\rfloor$ denotes the integer part. Thus in this case
$E_N$ depends on $\bde$ only through the combination $\bde+\bde^{\mathrm R}$, where
$\bde^{\mathrm R}$ is the reversed motif with components $\de^{\mathrm R}_j=\de_{N-j}$.
Conversely, if the energy function is linear and verifies
\begin{equation}
  \label{trinvrel}
  E_N(\bde)=f_N(\bde+\bde^{\mathrm R})
\end{equation}
for some function $f_N$, then the corresponding dispersion relation $\vep_N$ satisfies~\eqref{tim}.
Indeed, it suffices to impose~\eqref{trinvrel} for motifs of the form $(0,\dots,0,1,0,\dots,0)$,
which are allowed when $m>0$, or of the form $(1,\dots,1,0,1,\dots,1)$ when $n>0$. Motivated by
these observations, we shall say that a general Yangian-invariant spin model is
\emph{translationally invariant} whenever its energy function (not necessarily linear) satisfies
the identity~\eqref{trinvrel}. The purpose of this section is to study the minimum average
degeneracy of such models. 

By Eq.~\eqref{trinvrel}, the minimum average degeneracy of a translationally (and Yangian)
invariant spin model is given by
\begin{equation}\label{dNsmin}
d_{N,\mathrm{min}}^{(\mathrm s)}=\frac{(m+n)^N}{\nu_N^{(\mathrm s)}(m|n)},
\end{equation}
where $\nu_N^{(\mathrm s)}(m|n)$ is the cardinal of the set
\begin{equation}
  \label{DeNsmn}
  \De^{(\mathrm s)}_N(m|n)=\big\{\bde+\bde^{\mathrm R}\mid\bde\in\De_N(m|n)\big\}\,.
\end{equation}
Given a motif $\bde\in\De_N(m|n)$, we define its associated \emph{half-motif} as the vector
\[
\hat\bde=\begin{cases}
  (\de_1+\de_{N-1},\de_2+\de_{N-2},\dots,\de_{(N-1)/2}+\de_{(N+1)/2})\,,&N\ \text{odd}\\[1mm]
  (\de_1+\de_{N-1},\de_2+\de_{N-2},\dots,\de_{(N-2)/2}+\de_{(N+2)/2},\de_{N/2})\,,\quad
  & N\ \text{even}\,.
  \end{cases}
\]
Clearly, each vector $\bde+\bde^{\mathrm R}$ is uniquely determined by its corresponding
half-motif $\hat\bde$, so that in order to evaluate $\nu_N^{(\mathrm s)}(m|n)$ it suffices to
count the number of distinct half-motifs. It is also important to observe that all the components
of $\hat\bde$ can take the values $0,1,2$ except $\hat\de_{N/2}=\de_{N/2}$ (when $N$ is even),
which can only take the values $0$ or $1$.

In the genuinely supersymmetric case $\su(m|n)$ with $mn\ne0$, it is straightforward to compute
the number of distinct half-motifs, since in this case the motifs $\bde$ are arbitrary elements of
$\{0,1\}^{N-1}$. Thus, when $N$ is odd the half-motif $\hat\bde$ is an arbitrary element of
$\{0,1,2\}^{(N-1)/2}$, and therefore
\begin{equation}\label{nusNsupero}
\nu_N^{(\mathrm s)}(m|n)=3^{(N-1)/2}\,;\qquad mn\ne 0,\quad N\ \text{odd}.
\end{equation}
On the other hand, when $N$ is even we have
\begin{equation}\label{nusNsupere}
\nu_N^{(\mathrm s)}(m|n)=2\cdot 3^{(N-2)/2}\,;\qquad mn\ne 0,\quad N\ \text{even},
\end{equation}
since in this case the last component of $\hat\bde$ can only take the values $0$ or $1$. Hence the
minimum average degeneracy of a translationally (and Yangian) invariant spin model in the
genuinely supersymmetric case is given by
\begin{equation}\label{dNminTI}
d_{N,\mathrm{min}}^{(\mathrm s)}=
\begin{cases}
  \sqrt 3\,\left(\dfrac{m+n}{\sqrt 3}\right)^{N}\,,&N\ \text{odd}\\[3mm]
  \dfrac32\left(\dfrac{m+n}{\sqrt 3}\right)^{N}\,,&N\ \text{even}.
\end{cases}
\end{equation}
From the previous discussion it is obvious that $d_{N,\mathrm{min}}^{(\mathrm s)}$ exactly coincides with the
average degeneracy $d_N$ when there is no accidental degeneracy, namely when the function $f_N$ in
Eq.~\eqref{trinvrel} is injective.
\begin{example}
Consider the $\su(m|n)$ elliptic model~\eqref{H} in the special case $m=n=1$, which is obviously
invariant under integer translations. As shown in Ref.~\cite{FG14JSTAT}, any spin chain of the
form
\begin{equation}\label{H11}
H=\sum_{i<j}h(i-j)\Big(1-S_{ij}^{(1|1)}\Big)\,,
\end{equation}
where $h$ is an even $N$-periodic function, is isospectral to an $\su(1|1)$
Yangian-invariant spin model with linear energy function and dispersion relation
\begin{equation}\label{cENjh}
\vep_N(j)=\sum_{l=1}^{N-1}\Big(1-\cos(2\pi jl/N)\Big)\,h(l)\,.
\end{equation}
Since $\vep_N(j)=\vep(N-j)$, the dispersion relation~\eqref{cENjh} is also translationally
invariant in the sense of the definition~\eqref{trinvrel}. In particular, for the $\su(1|1)$
elliptic model~\eqref{H} the dispersion relation can be computed in closed form
using the techniques in Ref.~\cite{FG14JSTAT}, with the result
\begin{equation}\label{cENj11}
  \vep_N(j)=\frac{N}2\,\bigg(1-\frac EK\bigg)+\frac{N^2}{4K^2}\,\hat\eta_1-\frac1{16K'^2}\,
  \bigg[\wp(j/N)-\bigg(\ze(j/N)-2\eta_1\frac jN\bigg)^2\bigg]\,,\qquad j=1,\dots,N-1
\end{equation}
(and $\vep_N(0)=0$). In the previous formula $K'=K\Bigl(\sqrt{1-k^2}\,\Bigr)$,
\[
E(k)=\int_0^{\pi/2}\sqrt{1-k^2\sin^2\vp}\,\diff\vp
\]
is the complete elliptic integral of the second kind with modulus $k$, $\wp$ and $\ze$ denote the
Weierstrass elliptic functions with period lattice generated by $1$ and $\iu K/(NK')$,
$\eta_1=\ze(1/2)$ and $\hat\eta_1=\ze\big(1/2;1/2,\iu NK'/(2K)\big)$. Using Eqs.~\eqref{linfunc}
and~\eqref{cENj11} we have computed the spectrum of the $\su(1|1)$ elliptic chain~\eqref{H} for
$N\le 25$ and several values of the modulus $k$. In all cases, the average degeneracy $d_N$
coincides with the minimum average degeneracy~\eqref{dNminTI} of a translationally (and
Yangian) invariant spin model (cf.~Fig.~\ref{fig.avdegsu11}). Thus the
spectrum of the $\su(1|1)$ elliptic chain does not exhibit any accidental degeneracy.
\begin{figure}[h]
    \centering
    \includegraphics[width=.6\columnwidth]{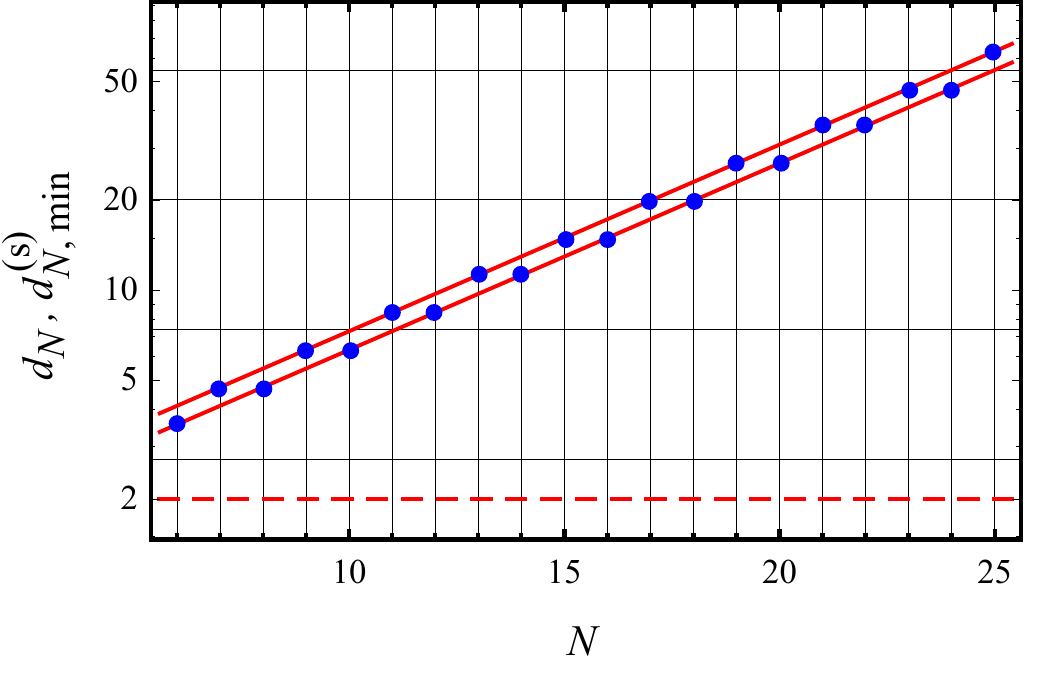}
    \caption{logarithmic plot of the average density of the $\su(1|1)$ elliptic chain with
      $k^2=1/2$ (blue dots), compared to the minimum average degeneracy~\eqref{dNminTI} of a
      translationally (and Yangian) invariant $\su(1|1)$ spin model (continuous red lines, the top
      one for odd $N$ and the bottom one for even $N$). The dashed horizontal line represents the
      minimum average degeneracy~\eqref{sumnbd} of a generic Yangian-invariant $\su(1|1)$ model.}
    \label{fig.avdegsu11}
  \end{figure}
\end{example}

In the non-supersymmetric (purely bosonic or fermionic) case, the computation of
$d_{N,\mathrm{min}}^{(\mathrm s)}$ is considerably more involved, due to the fact that not all
elements of $\{0,1\}^{N-1}$ are valid motifs. For this reason, we shall restrict ourselves in the
rest of this section to the simplest and most important $\su(2)$ case. In the bosonic case (which,
as we know, is equivalent to the fermionic one), an $\su(2)$ motif $\bde\in\{0,1\}^{N-1}$ cannot
contain two consecutive $1$'s. Hence, for odd $N$ the corresponding half-motif
$\hat\bde\in\{0,1,2\}^{(N-1)/2}$ satisfies the following two constraints:
\begin{description}\itemsep-3pt
\item[\hphantom{i}\rm i)] No $2$ can be preceded or followed by a $1$ or a $2$.
\item[\rm ii)] The last component of $\hat\bde$ cannot be equal to $2$.
\end{description}
It is also easy to see that any element of $\{0,1,2\}^{(N-1)/2}$ satisfying these conditions is a
valid half-motif $\hat\bde$ when $N$ is odd. On the other hand, when $N$ is even the last
component of a half-motif is the middle component of the corresponding motif, and therefore
condition ii) above should be replaced by the more stringent one
\begin{description}\itemsep-3pt
\item[\rm ii${}'$)] The half-motif $\hat\bde$ can only end in $0$ or $01$.
\end{description}
As before, it can be shown that any element of $\{0,1,2\}^{N/2}$ satisfying conditions i) and
ii${}'$) is a valid half-motif. Thus, in order to compute the minimum average degeneracy for odd
(respectively even) $N$ we need only count the number of elements of $\{0,1,2\}^{(N-1)/2}$
(respectively $\{0,1,2\}^{N/2}$) satisfying conditions i) and ii) (respectively i) and ii${}'$)).

Consider, to begin with, the case of odd $N=2r+1$, and denote by $\mu_r$ the number of vectors
$\hat\bde\in\{0,1,2\}^r$ satisfying the constraints i) and ii). Clearly, any such vector
$\hat\bde$ can only end in $x1$, $x\ms0$ and $20$, where $x$ is either $0$ or $1$. In the first
two cases, the first $r-1$ components of $\hat\bde$ still satisfy conditions i) and ii), so that
the number of valid half-motifs of length $r$ ending in $x1$ or $x\ms0$ is equal to $2\mu_{r-1}$.
On the other hand, if $\hat\bde$ ends in $20$ its first $r-2$ components are a vector belonging to
the set of half-motifs of length $r-2$ not ending in $1$, whose cardinal is obviously
$\mu_{r-2}-\mu_{r-3}$. We thus arrive at the recursion relation
\begin{subequations}\label{nur}
  \begin{equation}
    \label{rrTI}
    \mu_r=2\mu_{r-1}+\mu_{r-2}-\mu_{r-3}\,.
  \end{equation}
  It is straightforward to check that $\mu_1=2$, $\mu_2=5$ and $\mu_3=11$, which, together
  with~\eqref{rrTI}, completely determines $\mu_r$ for all $r$. In fact, it is immediate to verify
  that the latter initial conditions are equivalent to the simpler ones
  \begin{equation}
    \mu_{-2}=\mu_{-1}=0\,,\quad \mu_0=1\,.
    \label{incondodd}
  \end{equation}
\end{subequations}
Consider next the case of even $N=2r$, and let $\tmu_r$ denote the number of vectors
$\hat\bde\in\{0,1,2\}^r$ satisfying the constraints i) and ii${}'$). The reader should easily
convince himself that the valid motifs are precisely those from the $N=2r+1$ case not ending in
$11$, so that
\begin{equation}
  \tmu_r=\mu_r-\mu_{r-2}\,.
  \label{munu}
\end{equation}
Since $\mu_r$ (and hence $\mu_{r-2}$) is a solution of the linear recursion relation with constant
coefficients~\eqref{rrTI}, it follows that $\tmu_r$ satisfies the same recursion relation:
\begin{subequations}\label{mur}
  \begin{equation}
    \label{rrTIe}
    \tmu_r=2\tmu_{r-1}+\tmu_{r-2}-\tmu_{r-3}\,.
  \end{equation}
  However, the initial conditions are now $\tmu_1=2$, $\tmu_2=4$, $\tmu_3=9$, or equivalently
  \begin{equation}
    \label{incondeven}
    \tmu_{-2}=\tmu_{-1}=\tmu_0=1\,.
  \end{equation}
\end{subequations}
Thus, the minimum average degeneracy of an $\su(2)$ translationally (and Yangian) invariant spin
model is given by
\begin{equation}\label{dNminsu2}
d_{N,\mathrm{min}}^{(\mathrm s)}=\frac{2^N}{\nu_N^{(\mathrm s)}(2)}\,,
\end{equation}
where
\begin{equation}
  \label{su2smotifs}
  \nu_N^{(\mathrm s)}(2)=
  \begin{cases}
    \mu_{(N-1)/2}\,,\quad& N \text{ odd}\\[1mm]
    \tmu_{N/2}\,,\quad& N \text{ even.}
  \end{cases}
\end{equation}

The characteristic polynomial $p(\la)=\la^3-2\la^2-\la+1$ of the recursion
relation~\eqref{rrTI}-\eqref{rrTIe} has three real roots given by
\[
\la_k=\frac23\Big(1+\sqrt7\cos(\vp+2k\pi/3)\Big)\,,
\qquad k=0,1,2\,,
\]
with
\[
\vp\equiv\frac13\,\arctan\bigl(3\sqrt3\bigr)\,.
\]
As explained in
Section~\ref{abmad}, the sequences $\mu_r$ and $\tmu_r$ ($r=1,2,\dots$) can be expressed as
\[
\mu_r=\sum_{k=0}^2a_k\la_k^r\,,\qquad \tmu_r=\sum_{k=0}^2\ta_k\la_k^r\,,
\]
where the coefficients $a_k$ and $\ta_k$ are respectively determined by the initial
conditions~\eqref{incondodd} and \eqref{incondeven}. (In fact, from Eq.~\eqref{munu} it follows
that $\ta_k=a_k\big(1-\la_k^{-2}\big)$.) Since the root of $p(\la)$ with largest absolute value is
$\la_0\simeq2.24698$ (while $|\la_{1,2}|<1$), the asymptotic behavior of $\mu_r$ and $\tmu_r$ as
$r\to\infty$ is given by
\begin{equation}
\mu_r\underset{r\to\infty}{\simeq}a_0\,\la_0^r\,,\qquad
\tmu_r\underset{r\to\infty}{\simeq}\ta_0\,\la_0^r\,.
\label{numuas}
\end{equation}
The coefficients $a_0$ and $\ta_0$ can be readily computed in closed form, with the result
\begin{equation}
  a_0=\frac{4\big(1+\sqrt7\,\cos\vp\big)^2}{21(1+2\cos2\vp)}\simeq0.97869\,,\qquad
  \ta_0=\frac{4\big(1+\sqrt7\,\cos\vp\big)^2-9}{21(1+2\cos2\vp)}\simeq0.78485\,.
\end{equation}
From Eqs.~\eqref{su2smotifs} and~\eqref{numuas} we easily deduce the asymptotic behavior of the
minimum average degeneracy $d_{N,\mathrm{min}}^{(\mathrm s)}$ as $N\to\infty$ in the $\su(2)$
case, namely
\begin{equation}
  \label{dNmins2}
  d_{N,\mathrm{min}}^{(\mathrm s)}\underset{N\to\infty}{\simeq}\ga_{2,N}^{(\mathrm s)}\,\bigg(\frac{2}{\sqrt\la_0}\bigg)^N\,,
\end{equation}
where the coefficient $\ga_{2,N}^{(\mathrm s)}$ is given by
\begin{equation}
  \label{ga2s}
  \ga_{2,N}^{(\mathrm s)}=
  \begin{cases}
    \dfrac{\sqrt{\la_0}}{a_0}\,,\quad& N\text{ odd}\\[3mm]
    \dfrac{1}{\ta_0}\,,\quad& N\text{ even.}
  \end{cases}
\end{equation}
As $N\to\infty$, this minimum average degeneracy becomes considerably larger than the
corresponding one for a generic Yangian-invariant $\su(2)$ model. Indeed, from
  Eqs.~\eqref{dNmin2} and~\eqref{dNmins2} we obtain
\[
\frac{d_{N,\mathrm{min}}^{(\mathrm s)}}{d_{N,\mathrm{min}}}\underset{N\to\infty}\simeq
\al_N\Bigg[\frac{2}{\big(\sqrt5-1\big)\sqrt{\la_0}}\,\Bigg]^N\simeq \al_N\times1.07941^N\,,
\]
where $\al_N=2\ga_{2,N}^{(\mathrm s)}/\big(5-\sqrt 5\big)$ is close to $1$ ($1.10830$ for odd $N$
and $0.92197$ for even $N$).

\section{Average degeneracy of spin chains of Haldane--Shastry type}\label{sec.avdeg}

Ever since Haldane's original paper~\cite{Ha88}, it has been noted that the spectrum of the HS
chain exhibits a high degree of degeneracy, which is generally attributed to the Yangian symmetry
of this model. In fact, since both the $\su(m|n)$ HS and PF chains are Yangian-invariant spin
models, their average degeneracy is bounded below by $d_{N,\mathrm{min}}$ in Eqs.~\eqref{sumnbd}
(in the genuinely supersymmetric case) or~\eqref{MAD} (in the purely bosonic or fermionic case).
In the case of the FI chain, although its symmetry under the Yangian has not been proved
rigorously, since the spectrum is given by Eqs.~\eqref{HSPFspec} and \eqref{FI} the same bounds
apply. Roughly speaking, for the simplest $\su(2)$ case it can be checked that the average
degeneracy of the HS, PF and FI chains is of the same order of magnitude as the above bounds
---or, in the case of the HS chain, their refinement~\eqref{dNminsu2}-\eqref{su2smotifs} for
translationally invariant spin models--- {\em only} for low values of $N$ (say, $N\lesssim 10$).
In this section we shall study the average degeneracy of these models for much higher values of
$N$ (up to $N=50$), and show that, contrary to widespread belief, it cannot be solely explained by
their Yangian invariance.

\begin{figure}[h]
    \centering
    \includegraphics[width=.475\columnwidth]{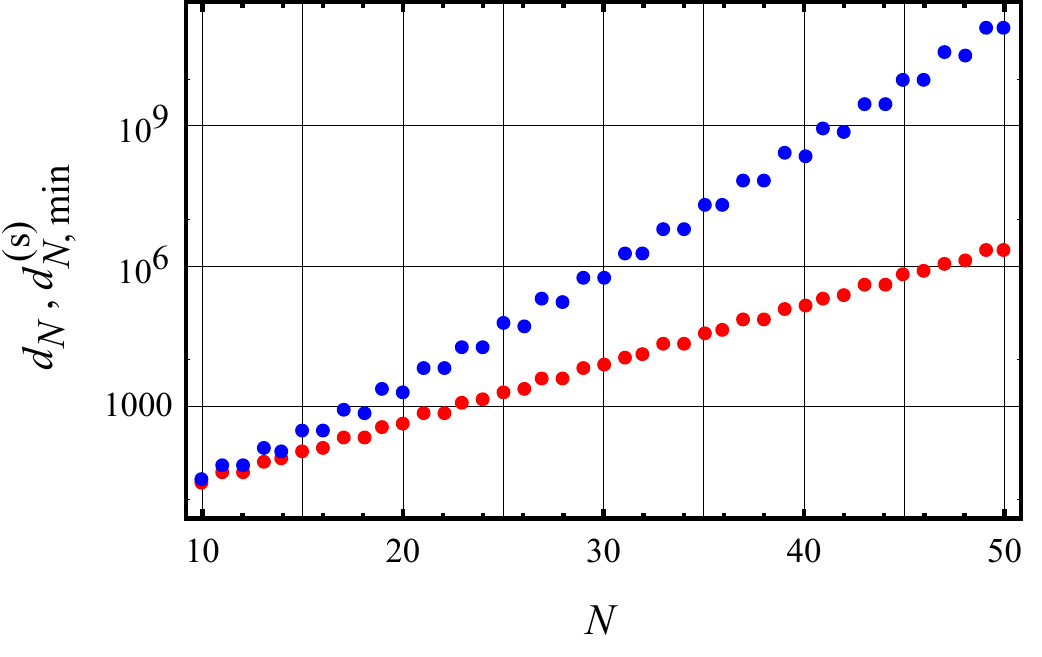}\hfill
    \includegraphics[width=.475\columnwidth]{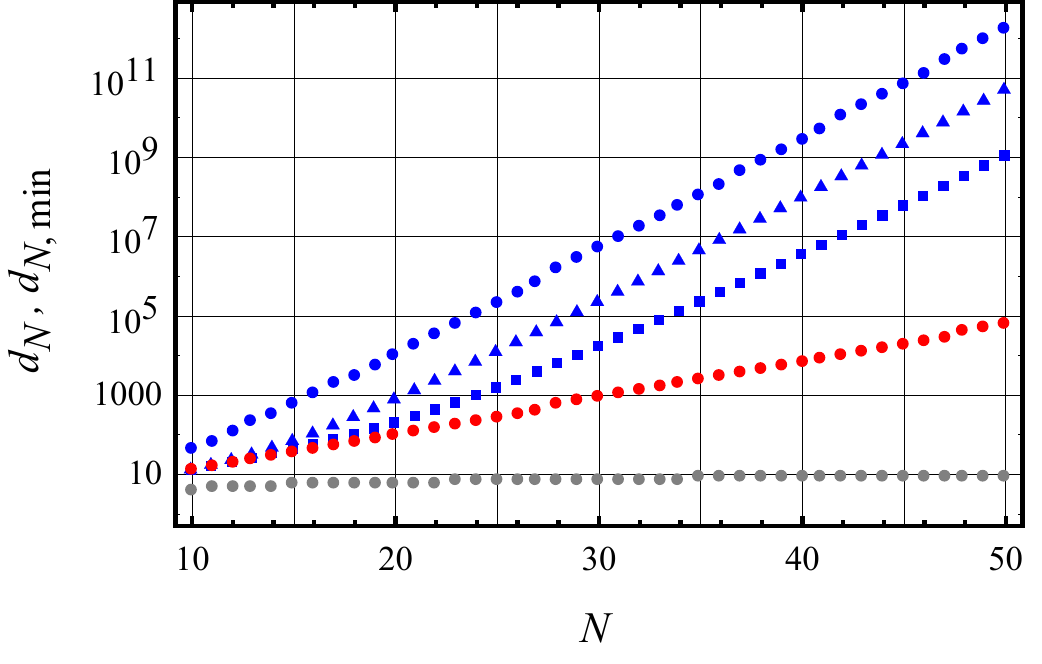}
    \caption{Left: logarithmic plot of the average degeneracy of the Haldane--Shastry $\su(2)$
      spin chain (blue circles), compared to the minimum average degeneracy of a translationally
      (and Yangian) invariant spin model (red circles). Right: analogous plot for the $\su(2)$
      Polychronakos--Frahm (blue circles) and Frahm--Inozemtsev (blue triangles for $\al=3$, blue
      squares for $\al$ irrational) spin chains, versus the minimum average degeneracy of a
      generic Yangian-invariant $\su(2)$ spin model (red circles) and of an $\su(2)$-invariant
      spin chain of the form~\eqref{HS-PF} (gray circles).}
    \label{fig.HSdegs}
\end{figure}

As explained above, for simplicity's sake we shall restrict ourselves to the $\su(2)$ (bosonic or
fermionic) case. For the PF chain, the computation of the average degeneracy is completely
straightforward, since its spectrum is a set of consecutive integers~\cite{Po94} and the minimum
and maximum energies can be found without difficulty. Indeed, consider (for instance) the bosonic
case. The minimum energy is clearly zero, corresponding to the $\su(2)$ motif $\bde=(0,\dots,0)$.
On the other hand, Eqs.~\eqref{cEN} and~\eqref{linfunc} imply that the maximum energy is obtained
from the motif $(\dots,0,1,0,1)$, with the result $E_{N,\mathrm{max}}=\big(N^2-\pi(N)\big)/4$.
Thus the number of distinct energy levels is given by
\begin{equation}\label{nuN2PF}
\nu_N(2)=E_{N,\mathrm{max}}+1=\frac{N^2-\pi(N)}4+1\qquad\text{(PF chain)}\,,
\end{equation}
and the average degeneracy is of course $d_N=2^N/\nu_N(2)$. In the case of the HS or FI chains, no
simple way of expressing in closed form the number of distinct levels as a function of $N$ is
known. For both of these chains, the most efficient way of computing the spectrum for relatively
high values of $N$ is to evaluate the partition function~\cite{FG05,BFGR10} in a recursive way.
More precisely, it can be shown that the partition function $Z_N$ of the (fermionic) FI chain is
determined by the recursion relation
\[
Z_{N+1}(q)=2q^{N(\al+N-1)}Z_N(q)+(1-q^{N(\al+N-1)})q^{(N-1)(\al+N-2)}Z_{N-1}(q),\qquad q\equiv\e^{-1/(k_{\mathrm B}T)}\,,
\]
with the initial conditions $Z_{-1}(q)=0$, $Z_0(q)=1$. The situation is a bit more involved for
the HS chain, since its dispersion relation~\eqref{cEN} depends explicitly on $N$. In this case,
the (fermionic) partition function is given by $Z_N(q)=Q_N(q;N)$, where $Q_k(q;N)$ is obtained from the
recursion relation
\[
Q_{k+1}(q;N)=2q^{k(N-k)}Q_k(q;N)+(1-q^{k(N-k)})q^{(k-1)(N-k+1)}Q_{k-1}(q;N),
\]
with the initial conditions $Q_{-1}(q;N)=0$, $Q_0(q;N)=1$. Proceeding in this way it is feasible
to compute the partition functions of the $\su(2)$ FI and HS chains for $N\le 50$ in a standard
desktop computer, and thus determine the number of distinct energies and the corresponding average
degeneracy $d_N$. In Figure~\ref{fig.HSdegs} we present logarithmic plots of $d_N$ for all three
chains of Haldane--Shastry type with $N$ ranging from $10$ to $50$, compared to its lower
bounds~\eqref{MAD} (for the PF and FI chains) and~\eqref{dNminsu2}-\eqref{su2smotifs} (for the HS
chain). It is apparent from this figure that as $N$ becomes moderately large (of the order of
$25$) $d_N$ grows much faster than expected for a generic Yangian-invariant spin model. The
ultimate reason for this behavior is the fact that these chains have a linear energy function
whose dispersion relation is a low degree polynomial. Indeed, we shall show that this implies that
the number of distinct levels has at most {\em polynomial} growth with $N$, compared to the
typical {\em exponential} growth for a generic Yangian-invariant spin model;
cf.~Eqs.~\eqref{nuNsuper}, \eqref{nuF}-\eqref{Fmnasymp}, \eqref{nusNsupero}-\eqref{nusNsupere},
\eqref{su2smotifs}-\eqref{numuas}. As a matter of fact, we have already proved that the number of
distinct levels of the $\su(2)$ PF chain is a second degree polynomial in~$N$; see
Eq.~\eqref{nuN2PF}. We shall next prove an analogous result for the $\su(2)$ HS and FI chains.

\begin{figure}[h]
    \centering
    \includegraphics[width=.6\columnwidth]{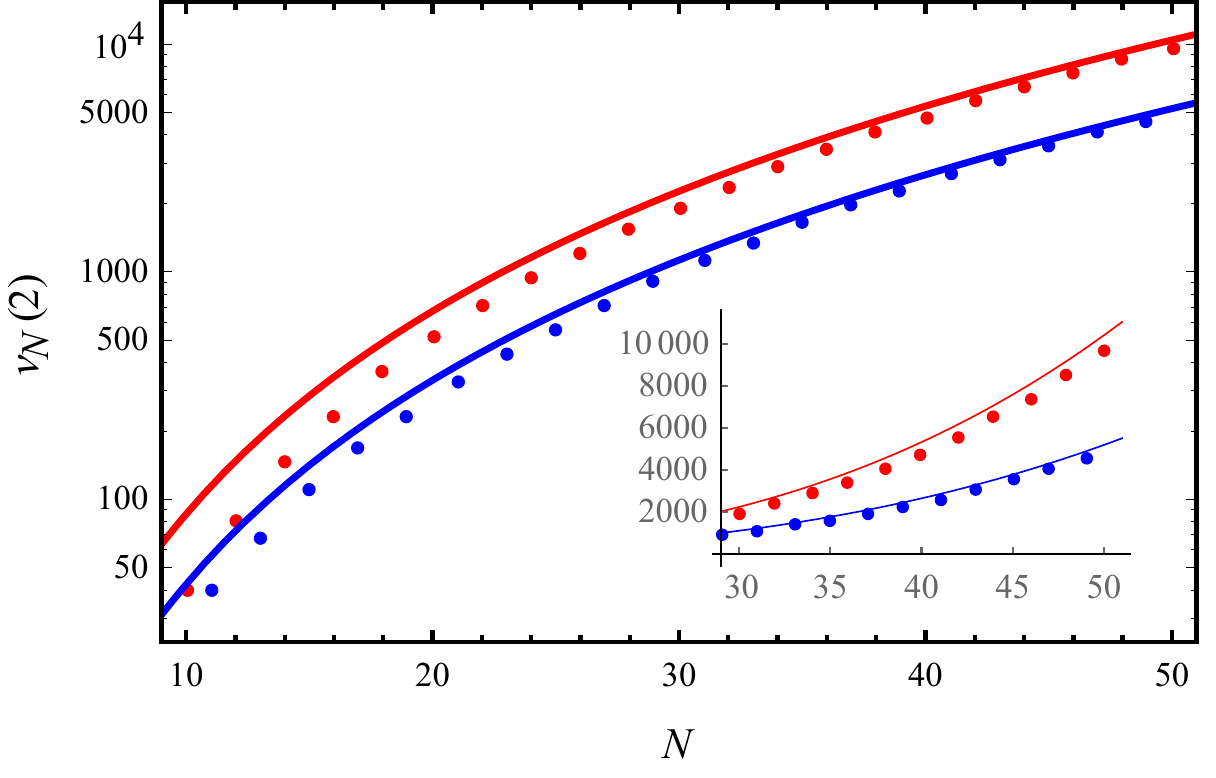}
    \caption{logarithmic plot of the number of distinct energy levels of the $\su(2)$
      Haldane--Shastry chain (red circles for even $N$, blue circles for odd $N$), compared to the
      upper bounds~\eqref{nuN2HS} (solid lines). Inset: same plot for the range $30\le N\le 50$,
      in ordinary (non-logarithmic) scale.}
    \label{fig.nuN2HS}
  \end{figure}

In the case of the HS chain, since the dispersion relation $\vep_N(j)$ is a polynomial in $j$ with
integer coefficients, and $\vep_N(j)>0$ for $1\le j\le N-1$, the energies are clearly nonnegative
integers. Moreover, from Eq.~\eqref{cEN} it follows that all the energies are even when $N$ is
odd. Since the minimum energy in the bosonic case is zero, the number of distinct energies
satisfies the inequality
\[
\nu_N(2)\le\frac{E_{N,\mathrm{max}}}{1+\pi(N)}+1\,,
\]
where $E_{N,\mathrm{max}}$ is the maximum energy of the bosonic chain. By
Eqs.~\eqref{Efb}-\eqref{Ezero}, this maximum energy is given by
\[
E_{N,\mathrm{max}}=\sum_{j=1}^{N-1}j(N-j)-E_{N,\mathrm{min}}^{\mathrm{(F)}}=\frac16\,N(N^2-1)
-E_{N,\mathrm{min}}^{\mathrm{(F)}}\,,
\]
where
\[
E_{N,\mathrm{min}}^{\mathrm{(F)}}=\frac1{12}\,N\big(N^2-4+3\pi(N)\big)\,.
\]
is the minimum energy of the fermionic chain computed in Ref.~\cite{BFGR08epl}. We thus obtain
\begin{equation}
  \label{nuN2HS}
  \nu_N(2)\le
  \begin{cases}
    \dfrac{N}{12}\,(N^2+2)+1\,,& N\text{ even}\\[4mm]
    \dfrac{N}{24}\,(N^2-1)+1\,,\quad& N\text{ odd}
  \end{cases}
\qquad\qquad(\text{HS chain})\,.
\end{equation}
In fact, as can be seen from Fig.~\ref{fig.nuN2HS}, these bounds provide a good approximation to
$\nu_N(2)$ for large~$N$.

Finally, in the case of the FI chain, the behavior of $\nu_N(2)$ is quite different when $\al$ is
rational or irrational. Indeed, if $\al=a/b$ is an irreducible rational number then $b E_N(\bde)$
is a nonnegative integer. Thus the number of distinct energies is bounded above by
$bE_{N,\mathrm{max}}+1$, where $E_{N,\mathrm{max}}$ is again the maximum energy of the bosonic
chain. Proceeding as before and using the value of $E_{N,\mathrm{min}}^{\mathrm{(F)}}$ computed in
Ref.~\cite{BFGR10} we readily obtain
\begin{equation}
  \label{nuN2FI}
  \nu_N(2)\le
  \begin{cases}
    \dfrac{N}{12}\,\big(2bN^2+3(a-b)N-2b\big)+1\,,& N\text{ even}\\[4mm]
    \dfrac{1}{12}\,(N^2-1)\big(2bN+3(a-b)\big)+1\,,\quad& N\text{ odd}
  \end{cases}
\qquad\qquad(\text{FI chain},\ \al\in\QQ)\,.
\end{equation}
On the other hand, when $\al$ is irrational we can write
\[
E_N(\bde)=\al\sum_{j=1}^{N-1}j\de_j+\sum_{j=1}^{N-1}j(j-1)\de_j\equiv\al E_N^{(0)}(\bde)
+E_N^{(1)}(\bde)\,,
\]
where the functions $E_N^{(i)}(\bde)$ ($i=0,1$) take nonnegative integer values and (in the
bosonic case) actually vanish when $\bde$ is the zero motif. Hence
\[
\nu_N(2)\le\big(E_{N,\mathrm{max}}^{(0)}+1\big)\big(E_{N,\mathrm{max}}^{(1)}+1\big)\,,
\]
where $E_{N,\mathrm{max}}^{(i)}$ is the maximum value of $ E_N^{(i)}(\bde)$ over all valid
$\su(2)$ bosonic motifs $\bde$. Although $E_{N,\mathrm{max}}^{(i)}$ can be computed without
difficulty, for our purposes it suffices to note that
\[
E_{N,\mathrm{max}}^{(0)}\le\sum_{j=1}^{N-1}j=\frac N2\,(N-1)\,,\qquad
E_{N,\mathrm{max}}^{(0)}\le\sum_{j=1}^{N-1}j(j-1)=\frac N3\,(N-1)(N-2)\,,
\]
from which it easily follows that
\begin{equation}
  \label{nuN2FIi}
  \nu_N(2)\le \frac{N^5}{6}\qquad(\text{FI chain},\ \al\notin\QQ)\,.
\end{equation}

Although the previous discussion has been restricted to the $\su(2)$ Haldane--Shastry type chains,
its main conclusions extend to the whole class of $\su(m|n)$ Yangian-invariant spin models studied
in Refs.~\cite{BBH10,BB12}, whose energy function is of the form~\eqref{linfunc} with a dispersion relation
$\vep_N(j)$ \emph{polynomial} in $j$ and $N$. Indeed, suppose that
\begin{equation}
  \label{poldisp}
  \vep_N(j)=\sum_{(r,s)\in I}\al_{rs}N^r j^s\,,
\end{equation}
where $I$ is a finite subset of $\NN_0\times\NN_0$, $\NN_0$ being the set of nonnegative integers, and
$\al_{rs}$ is independent of $N$ and non-zero for all $(r,s)\in I$. The energy function is then
given by
\[
E_N(\bde)=\sum_{(r,s)\in I}\al_{rs}N^r \sum_{j=1}^{N-1}j^s\de_j\equiv
\sum_{(r,s)\in I}\al_{rs}N^rE_N^{(s)}(\bde)\,,
\]
where each function $E_N^{(s)}(\bde)$ takes nonnegative integer values and is bounded above by the
power sum
\begin{equation}\label{Sbound}
  S_N^{(s)}\equiv\sum_{j=1}^{N-1}j^s=\frac{N^{s+1}}{s+1}+\Or(N^{s})\,.
\end{equation}
We thus obtain the polynomial bound
\begin{equation}
  \label{nuNmnpoly}
  \nu_N(m|n)\le \prod_{(r,s)\in I}\big(S_N^{(s)}+1\big)=\prod_{s}\big(S_N^{(s)}+1\big)^{r(s)}\,,
\end{equation}
where $r(s)$ is the number of indices $r$ such that $(r,s)\in I$, and the last product ranges over
all $s$ such that $r(s)>0$. Note that, on account of Eq.~\eqref{Sbound}, the
bound~\eqref{nuNmnpoly} satisfies
\[
\prod_{s}\big(S_N^{(s)}+1\big)^{r(s)}=\Or\left(N^{\sum_s(s+1)r(s)}\right)\,.
\]

The bound in Eq.~\eqref{nuNmnpoly} can be considerably strengthened when all the coefficients
$\al_{rs}$ in the dispersion relation~\eqref{nuNmnpoly} are \emph{rational} numbers, as is the case
with the PF and HS chains (or the FI chain, when $\al\in\QQ$). Indeed, assume that
$\al_{rs}=a_{rs}/b$, where $a_{rs}\in\ZZ$ and $b\in\NN$. In this case $bE_N(\bde)$ takes only
integer values, so that
\[
\nu_N(m|n)\le b\Big(E_{N,\mathrm{max}}-E_{N,\mathrm{min}}\Big)+1\,,
\]
where $E_{N,\mathrm{min}}$ and $E_{N,\mathrm{max}}$ denote respectively the minimum and maximum
energies. From the inequality
\[
\big|\vep_N(j)\big|\le\sum_{(r,s)\in I}|\al_{rs}|N^rj^s\le N^k\sum_{(r,s)\in I}|\al_{rs}|\equiv A
N^k\,,
\qquad 1\le j\le N-1\,,
\]
where $k=\max\limits_{(r,s)\in I}(r+s)$ is the total degree of $\vep_N(j)$ as a polynomial in
$(j,N)$, we immediately obtain
\[
E_{N,\mathrm{max}}-E_{N,\mathrm{min}}\le 2\max_{\bde\in\De_N(m|n)}\big|E_N(\bde)\big|\le2N\max_{1\le
  j\le N-1}\big|\vep_N(j)\big|\le 2AN^{k+1}\,,
\]
and hence
\begin{equation}
  \label{nubdint}
  \nu_N(m|n)\le (2bA+1)N^{k+1}\,.
\end{equation}
Note that this equation is in agreement with the asymptotic behavior of the 
bounds~\eqref{nuN2PF}--\eqref{nuN2FI} for the PF, HS and FI chains (with $\al\in\QQ$ in the latter
case).

\section{Anyons and Fibonacci numbers}\label{sec.anyons}

The elementary excitations of certain quantum many-body systems can be described in terms of
\emph{anyons}, i.e., effective quasi-particles obeying neither the Bose--Einstein nor the
Fermi--Dirac statistics~\cite{LM77,Wi82,Wi82b}. The original definition of anyons is intrinsically
two-dimensional, since it requires that the wave function change by a suitable phase under
exchange of two identical particles, which is only consistent in two dimensions. An alternative
definition of anyons in terms of a generalized exclusion principle was proposed by
Haldane~\cite{Ha91b}. According to this definition, in a system of (identical, non-interacting)
anyons the number of available one-particle states (or \emph{orbitals}) decreases by $g\De k$ when
the number of particles increases by $\De k$. The constant $g$, which must be a rational number,
is called Haldane's \emph{exclusion parameter}. Clearly, bosons and fermions have exclusion
parameter $g=0$ and $g=1$, respectively. The advantage of Haldane's definition is that, in
contrast with the original one, it is also valid in one dimension. For instance, it is well
known that when the coupling constant is a nonnegative integer $a$ the spectrum of the
$N$-particle Calogero model~\cite{Ca71} is that of a system of $N$ free anyons with exclusion
parameter $g=a$ in a harmonic oscillator potential~(see, e.g., the review paper~\cite{Po06}).

Another important example of one-dimensional anyons is furnished by the $\su(2)$ Haldane--Shastry
chain, whose ``spinon'' excitations (in the antiferromagnetic case) are known to be anyons with
$g=1/2$~\cite{Ha91b}. It should be clear that this property is in fact shared by \emph{all}
$\su(2)$ (fermionic or bosonic) Yangian-invariant spin models, regardless of their energy
function. To see this, consider an $\su(2)$ (say, bosonic) motif $\bde$ and its associated border
strip $\bk=(k_1,\dots,k_r)$, where $k_1+\cdots+k_r=N$ is the number of spins. In the $\su(2)$
case, it is a simple matter to compute the dimension of the corresponding $Y(\gl(2|0))$-module
$V_{\bk}(2|0)\equiv V_\bk$. Indeed, a (bosonic) $\su(2)$ border strip $\bk$ cannot contain rows
with more than two boxes. Moreover, any tableau corresponding to $\bk$ must necessarily contain
the sequence $01$ (from left to right) in any row with two boxes. The dimension of $V_{\bk}$ is
thus given by the number of ways of filling the remaining (``empty'') boxes in each column of the
border strip $\bk$ (from top to bottom) with sequences of the form $0\cdots01\cdots1$ (where the
number of 0's or 1's can of course be zero). Thus the $r$ columns of $\bk$ play the role of
available orbitals, and its $N-2r+2$ empty boxes that of particles with two internal degrees of
freedom (0 or 1). (These particles are obviously bosons, since each orbital can be occupied by
more than one particle with the same quantum numbers.) Thus in this case we have $k=N-2r+2$ and
$\De r=-\De k/2$, so that Haldane's definition is satisfied with $g=1/2$. This shows that
\emph{any} $\su(2)$ Yangian-invariant model is indeed equivalent to a system of anyons with two
internal degrees of freedom and exclusion parameter $g=1/2$ (\emph{semions}, in Haldane's
terminology).

Haldane's definition of anyons can be equivalently formulated in terms of the total number of
states in a system with $k$ anyons and $\cN$ one-particle states (or \emph{orbitals}). Indeed, for
anyons with exclusion parameter $g$ this number is given by
\begin{equation}
  \label{numstates}
  w_k(g) = \binom{\cN-(g-1)(k-1)}{k}\,.
\end{equation}
When $g$ is a natural number, a very simple realization of this statistics can be obtained by
considering a system of $k$ particles and $\cN$ orbitals satisfying the following generalized
Pauli exclusion principle: two occupied orbitals must be separated by at least $g-1$ empty ones.
This requirement is clearly reminiscent of the rule satisfied by the bosonic $\su(2)$ motifs. More
precisely, we can interpret such a motif $\bde\in\De_N(2|0)$ as encoding a state of a quantum
system with $N-1$ orbitals (the components of $\bde$), with a 1 indicating an occupied orbital and
0 an empty one. With this interpretation, the 1's in the motif $\bde$ are quasi-particles
satisfying the generalized Pauli principle with $g=2$, and thus can be regarded as anyons with
exclusion parameter $g=2$. This point of view is particularly natural when the energy function is
linear (cf.~Eq.~\eqref{linfunc}), as is the case with spin chains of Haldane--Shastry type
(cf.~Eqs.~\eqref{HSPFspec}-\eqref{FI}). Indeed, for such Yangian-invariant models the energy
function can be written as
\begin{equation}\label{ENrap}
E_N(\bde)=\sum_{i=1}^{r-1}\vep_N(K_i)\,,
\end{equation}
where $K_i$ is the position of the $i$-th 1 in $\bde$. It should be stressed that this
interpretation is not equivalent to the one discussed above in terms of spinons, for which $g=1/2$.
Indeed, in the latter interpretation the 1's in the motif (or, more accurately, the sequences
$(0,1)$) simply separate the orbitals, while the remaining 0's correspond to the particles. In
particular, a fundamental difference between both interpretations is that in the new one the
degeneracy associated to each motif (that is, the dimension of the corresponding invariant module
$V_\bk$) is not taken into account. For this reason, the new interpretation defines a different
type of system, whose Hilbert space dimension equals the number $\nu_N(2)$ of distinct $\su(2)$
motifs. Since $\nu_N(2)$ is given by the Fibonacci number $F_{N+1}$ (cf.~Eq.~\eqref{nuF}), from
Eq.~\eqref{numstates} with $\cN=N-1$ and $g=2$ we obtain the remarkable identity
\begin{equation}
  \label{Fibid}
  F_{N+1}=\sum_{k=0}^{\lfloor N/2\rfloor}\binom{N-k}{k}\,,
\end{equation}
where the upper limit is determined by the condition $N-k\ge k$. In other words, the $(N+1)$-th
Fibonacci number is obtained by diagonally summing the binomial coefficients arranged in a Pascal
triangle~\cite{Mi60}. This classical number-theoretic result can thus be interpreted as expressing
the dimension of the Hilbert space of a system of $g=2$ anyons with $N-1$ orbitals in terms of the
corresponding dimensions of its $k$-particle subspaces.

Motivated by the previous discussion, we shall next consider the bosonic $\su(m)$ motifs with
$m>2$ as representing the states of a system of quasi-particles (namely, the 1's in the motif)
with $N-1$ orbitals (the motif's components), subject to the exclusion principle that there can be
no more than $m-1$ consecutive occupied orbitals. As before, this rule obviously does not take
into account the degeneracy associated to each motif, so that the system under consideration is
\emph{not} an $\su(m)$ Yangian-invariant spin model. It can instead be regarded as an effective
model, whose spectrum~\eqref{ENrap} reproduces the set of {\em distinct} energy levels of a
generic (bosonic) $\su(m)$ Yangian-invariant spin model with linear energy function and dispersion
relation $\vep_N(j)$.

We shall next compute the statistical weight $w_k$ for the new system just
introduced. To this end, let $\bde\in\De_N(m|0)$ be a bosonic $\su(m)$ motif with $j_0$ single 1's
(i.e., $(1,0)$ sequences), $j_1$ double 1's (i.e, $(1,1,0)$ sequences), \dots, and $j_{m-2}$
sequences $(\underbrace{1,\dots,1}_{m-1},0)$, with
\begin{equation}\label{jisk}
j_0+2j_1+\dots +(m-1)j_{m-2}=k\,.
\end{equation}
Removing the zero in each sequence $(1,0)$, $(1,0)$ in each sequence $(1,1,0)$, \dots,
$(\underbrace{1,\dots,1}_{m-2},0)$ in each sequence $(\underbrace{1,\dots,1}_{m-1},0)$, except for
the zero (if any) following the last 1 in the motif, we are left with a sequence
$\bar\bde\in\{0,1\}^{N-k}$ containing $j_0+\cdots +j_{m-2}$ $1$'s. Conversely, it is clear that
from \emph{any} such vector $\bar\bde\in\{0,1\}^{N-k}$ one can construct a bosonic $\su(m)$ motif
with $j_0$ single 1's, $j_1$ double 1's, \dots, and $j_{m-2}$ sequences
$(\underbrace{1,\dots,1}_{m-1},0)$ by adding a zero after $j_0$ 1's, the sequence $(1,0)$ after
$j_1$ of the remaining 1's, etc. Thus, the number $w(j_0,\dots,j_{m-2})$ of motifs with $j_0$
single 1's, $j_1$ double 1's, etc., is given by the number of ways of choosing the positions of
the $j_0$ 1's starting a sequence $(1,0)$, the $j_1$ 1's starting a sequence $(1,1,0)$, etc.,
among the integers $1,\dots,N-k$. In other words,
\[
w(j_0,\dots,j_{m-2})=\frac{(N-k)!}{(N-k-\sum\limits_{i=0}^{m-2}j_i)!\prod\limits_{i=0}^{m-2}j_i!}\,.
\]
The statistical weight $w_k$ is the sum of all $w(j_0,\dots,j_{m-2})$ whose
arguments~$(j_0,\dots,j_{m-2})$ satisfy~\eqref{jisk}, namely
\begin{equation}\label{wksum}
  w_k=\sum_{\substack{j_0,\dots,j_{m-2}\ge0\\j_0+2j_1+\dots +(m-1)j_{m-2}=k}}
  \frac{(N-k)!}{\Big(N-k-\sum\limits_{i=0}^{m-2}j_i\Big)!\prod\limits_{i=0}^{m-2}j_i!}\,.
\end{equation}
Note that for $m=2$ this equation obviously reduces to Eq.~\eqref{numstates} with $g=2$ and
$\cN=N-1$. Eliminating $j_0$ with the help of Eq.~\eqref{jisk} and explicitly enforcing the
condition $\sum_{i=0}^{m-2}j_i\le N-k$ arising from the first factor in the denominator we obtain
the equivalent expression
\begin{equation}\label{wksumfinal}
  w_k=\sum_{\substack{j_1,\dots,j_{m-2}\ge0\\2j_1+3j_2+\dots+(m-1)j_{m-2}\le k\\j_1+2j_2+\dots +(m-2)j_{m-2}\ge 2k-N}}
  \frac{(N-k)!}{\big(N-2k+\sum\limits_{i=1}^{m-2}ij_i\Big)!\,
    \Big(k-\sum\limits_{i=1}^{m-2}(i+1)j_{i}\Big)!\prod\limits_{i=1}^{m-2}j_i!}\,.
\end{equation}
For instance, for $m=3$ we have
\[
w_k=\sum_{j=\max(2k-N,0)}^{\lfloor k/2\rfloor}\frac{(N-k)!}{(N-2k+j)!(k-2j)!j!}\,,\qquad m=3\,.
\]
The total dimension of the system's Hilbert space, which is the same as the number $\nu_N(m)$ of
bosonic $\su(m)$ motifs for a Yangian-invariant spin model with $N$ spins, is obtained by summing
the statistical weight $w_k$ over all possible values of $k$. If $N-1=r m+ s$ with
$0\le s\le m-1$, the maximum value of $k$ is clearly
\[
r(m-1)+s=\frac{m-1}{m}\,(N-1-s)+s=\frac{m-1}{m}\,N-\frac{m-1-s}{m}
=\left\lfloor\frac{m-1}{m}\,N\right\rfloor\,,
\]
since $0\le m-1-s\le m-1$. From Eqs.~\eqref{wksumfinal} and~\eqref{nuF} we finally obtain the
identity
\begin{equation}\label{Fibidm}
  F^{(m)}_{N+m-1}=\sum_{k=0}^{\lfloor(m-1)N/m\rfloor}
   \sum_{\substack{j_1,\dots,j_{m-2}\ge0\\2j_1+3j_2+\dots+(m-1)j_{m-2}\le k\\j_1+2j_2+\dots +(m-2)j_{m-2}\ge 2k-N}}
  \frac{(N-k)!}{\big(N-2k+\sum\limits_{i=1}^{m-2}ij_i\Big)!\,
    \Big(k-\sum\limits_{i=1}^{m-2}(i+1)j_{i}\Big)!\prod\limits_{i=1}^{m-2}j_i!}\,.
\end{equation}
An alternative version of the latter equation follows by using Eq.~\eqref{wksum} for $w_k$ and
eliminating $k$ using Eq.~\eqref{jisk}, namely
\begin{equation}\label{Fibidm2}
  F^{(m)}_{N+m-1}=\sum_{\substack{j_0,\dots,j_{m-2}\ge0\\2j_0+3j_1+\dots+mj_{m-2}\le N}}\frac{\Big(N-\sum\limits_{i=0}^{m-2}(i+1)j_i\Big)!}{\Big(N-\sum\limits_{i=0}^{m-2}(i+2)j_i\Big)!\prod\limits_{i=0}^{m-2}j_i!}\,.
\end{equation}
This is essentially the classical combinatorial expression for the generalized Fibonacci numbers
proved in~\cite{Mi60} (cf.~Eq.~($28''$) in the latter reference).

It is apparent from Eqs.~\eqref{numstates} (with $\cN=N-1)$ and~\eqref{wksumfinal} that for $m>2$
the quasi-particles associated to the (bosonic) $\su(m)$ motifs of a Yangian-invariant spin model
are not anyons. Note, however, that there is a generalization of Haldane's
definition~\eqref{numstates}, due to Murthy and Shankar~\cite{MS94a}, to systems with an
infinite-dimensional one-particle Hilbert space. The latter generalization is essentially based on
the behavior of the statistical weight~\eqref{numstates} as the number of orbitals $\cN$ tends to
infinity for any fixed $k$, namely
\begin{equation}
  \label{wkasymp}
  w_k(g) = \frac{\cN^k}{k!}+\bigg(\frac12-g\bigg)\,\frac{\cN^{k-1}}{(k-2)!}+\Or\big(\cN^{k-2}\big)\,,
\end{equation}
from which it follows that
\begin{equation}
  \label{glim}
  \frac12-g=\lim_{\cN\to\infty}\frac{\cN}{k(k-1)}\bigg(\frac{k!\,w_k(g)}{\cN^k}-1\bigg)\,.
\end{equation}
This observation suggests looking at the asymptotic behavior of the density of
states~\eqref{wksumfinal} as $\cN=N-1\to\infty$ with $k$ fixed. To this end, note first of all
that $j_i\le k/(i+1)$ on account of the restriction $\sum_{i=1}^{m-2}(i+1)j_i\le k$, and that
\[
\frac{(\cN-k+1)!}{(\cN-2k+l+1)!}\underset{\cN\to\infty}=\cN^{k-l}+\Or\big(\cN^{k-l-1}\big)\,.
\]
Thus the leading terms in the sum~\eqref{wksumfinal} are obtained for $j_1=\dots=j_{m-2}=0$ and
$j_1=1$, $j_2=\dots=j_{m-2}=0$, or more precisely
\[
w_k=\frac{(\cN-k+1)!}{(\cN-2k+1)!k!}+\frac{(\cN-k+1)!}{(\cN-2k+2)!(k-2)!}+\Or\big(\cN^{k-2}\big)
=\binom{\cN-k+1}{k}+\frac{\cN^{k-1}}{(k-2)!}+\Or\big(\cN^{k-2}\big)\,.
\]
Using Eqs~\eqref{numstates}-\eqref{wkasymp} with $g=2$ to expand the first term on the RHS we
immediately obtain
\[
w_k=\frac{\cN^k}{k!}-\frac12\,\frac{\cN^{k-1}}{(k-2)!}+\Or\big(\cN^{k-2}\big)\,.
\]
Thus as $\cN\to\infty$ with $k$ fixed the density of states~\eqref{wksumfinal} satisfies
Eq.~\eqref{wkasymp} with $g=1$. Note that, as already observed by Murthy and Shankar, if the
system's maximum and minimum energies are independent of $\cN$ (or, more generally, the spectrum
is bounded by $\cN$-independent constants) the limit $\cN\to\infty$ with $k$ fixed can be
naturally interpreted as the continuum limit. We conclude that in the continuum limit the
(bosonic) $\su(m)$ motifs with $m>2$ behave as a system of \emph{fermions}, according to Murthy
and Shankar's definition.

\section{Conclusions}\label{sec.conc}

In this paper we consider a wide class of finite-dimensional Yangian-invariant quantum models
---which we have called Yangian-invariant spin models--- closely related to representations of the
Yangian $Y(\gl(m|n))$ labeled by border strips. This class includes several widely studied
systems, such as spin chains of Haldane--Shastry type and certain families of one-dimensional
vertex models. A common feature of all Yangian-invariant spin models is the relatively high
degeneracy of their spectrum, essentially due to their symmetry under the Yangian. As a
quantitative measure of the degree of degeneracy of these models, we compute in closed form the
intrinsic average degeneracy of any such model stemming from its Yangian invariance. It turns out
that, in the non-supersymmetric case, this minimum average degeneracy is expressed in terms of
generalized Fibonacci numbers. Using several fundamental properties of these numbers, we derive
the asymptotic behavior of the minimum average degeneracy as the number of spins tends to
infinity. Our results provide a stringent test for a quantum system to be a Yangian-invariant spin
model. As an example, we apply this test to the $\su(m|n)$ supersymmetric version of Inozemtsev
chain, which has been studied in the context of the gauge-string duality as a candidate for
reproducing the spectrum of the dilation operator at several loops; see, e.g., \cite{Be12}. We
have also refined the previous results on the minimum average degeneracy for models that are in
addition translationally invariant, both in the $\su(m|n)$-supersymmetric and in the simplest
$\su(2)$ cases.

The second main problem that we address in this paper is the analysis of the precise role played
by the Yangian symmetry in the high degeneracy of the spectra of spin chains of Haldane--Shastry
type. We show that these chains are much more degenerate than generic Yangian-invariant spin
models, since the number of distinct levels grows polynomially for the former systems and
exponentially for the latter. This accidental degeneracy is essentially due to the polynomial
character of the dispersion relation, and is therefore shared by the family of one-dimensional
vertex models studied in Refs.~\cite{BBH10,BB12}. Moreover, if the coefficients in the dispersion
relation are rational numbers (as is the case for the HS and PF chains), we show that the
accidental degeneracy is even higher, as illustrated by the FI chain with a rational
vs.~irrational parameter. Finally, we consider an effective model of quasi-particles describing
the distinct energy levels of a non-supersymmetric Yangian-invariant spin model. We show that in
the $\su(2)$ case these quasi-particles are anyons with exclusion parameter $g=2$, while in the
$\su(m)$ case with $m>2$ they become fermions in the continuum limit. We also provide a simple
interpretation of several combinatorial expressions for $m$-nacci numbers in terms of the
statistical weights of the effective model of $\su(m)$ quasi-particles.

\section*{Acknowledgments}

This work was supported in part by Spain's MINECO under grant no.~FIS2011-22566 and by the
Universidad Complutense and Banco Santander under grant no. GR3/14-910556.


\begin{thebibliography}{50}
\expandafter\ifx\csname natexlab\endcsname\relax\def\natexlab#1{#1}\fi
\providecommand{\bibinfo}[2]{#2}
\ifx\xfnm\relax \def\xfnm[#1]{\unskip,\space#1}\fi
\bibitem[{Kirillov et~al.(1997)Kirillov, Kuniba, and Nakanishi}]{KKN97}
\bibinfo{author}{A.~N. Kirillov}, \bibinfo{author}{A.~Kuniba},
  \bibinfo{author}{T.~Nakanishi}, \bibinfo{journal}{Commun. Math. Phys.}
  \bibinfo{volume}{185} (\bibinfo{year}{1997}) \bibinfo{pages}{441--465}.
\bibitem[{Nazarov and Tarasov(1998)}]{NT95}
\bibinfo{author}{M.~Nazarov}, \bibinfo{author}{V.~Tarasov},
  \bibinfo{journal}{J. Reine Angew. Math} \bibinfo{volume}{496}
  (\bibinfo{year}{1998}) \bibinfo{pages}{181--212}.
\bibitem[{Haldane et~al.(1992)Haldane, Ha, Talstra, Bernard, and
  Pasquier}]{HHTBP92}
\bibinfo{author}{F.~D.~M. Haldane}, \bibinfo{author}{Z.~N.~C. Ha},
  \bibinfo{author}{J.~C. Talstra}, \bibinfo{author}{D.~Bernard},
  \bibinfo{author}{V.~Pasquier}, \bibinfo{journal}{Phys. Rev. Lett.}
  \bibinfo{volume}{69} (\bibinfo{year}{1992}) \bibinfo{pages}{2021--2025}.
\bibitem[{Hikami(1995)}]{Hi95npb}
\bibinfo{author}{K.~Hikami}, \bibinfo{journal}{Nucl. Phys. B}
  \bibinfo{volume}{441} (\bibinfo{year}{1995}) \bibinfo{pages}{530--548}.
\bibitem[{Bouwknegt and Schoutens(1996)}]{BS96}
\bibinfo{author}{P.~Bouwknegt}, \bibinfo{author}{K.~Schoutens},
  \bibinfo{journal}{Nucl. Phys. B} \bibinfo{volume}{482} (\bibinfo{year}{1996})
  \bibinfo{pages}{345--372}.
\bibitem[{Haldane(1988)}]{Ha88}
\bibinfo{author}{F.~D.~M. Haldane}, \bibinfo{journal}{Phys. Rev. Lett.}
  \bibinfo{volume}{60} (\bibinfo{year}{1988}) \bibinfo{pages}{635--638}.
\bibitem[{Shastry(1988)}]{Sh88}
\bibinfo{author}{B.~S. Shastry}, \bibinfo{journal}{Phys. Rev. Lett.}
  \bibinfo{volume}{60} (\bibinfo{year}{1988}) \bibinfo{pages}{639--642}.
\bibitem[{Haldane(1994)}]{Ha93}
\bibinfo{author}{F.~D.~M. Haldane}, in: \bibinfo{editor}{A.~Okiji},
  \bibinfo{editor}{N.~Kawakami} (Eds.), \bibinfo{booktitle}{Correlation Effects
  in Low-dimensional Electron Systems}, volume \bibinfo{volume}{118} of
  \textit{\bibinfo{series}{Springer Series in Solid-state Sciences}}, pp.
  \bibinfo{pages}{3--20}.
\bibitem[{Polychronakos(1993)}]{Po93}
\bibinfo{author}{A.~P. Polychronakos}, \bibinfo{journal}{Phys. Rev. Lett.}
  \bibinfo{volume}{70} (\bibinfo{year}{1993}) \bibinfo{pages}{2329--2331}.
\bibitem[{Frahm(1993)}]{Fr93}
\bibinfo{author}{H.~Frahm}, \bibinfo{journal}{J. Phys. A: Math. Gen.}
  \bibinfo{volume}{26} (\bibinfo{year}{1993}) \bibinfo{pages}{L473--L479}.
\bibitem[{Hikami and Basu-Mallick(2000)}]{HB00}
\bibinfo{author}{K.~Hikami}, \bibinfo{author}{B.~Basu-Mallick},
  \bibinfo{journal}{Nucl. Phys. B} \bibinfo{volume}{566} (\bibinfo{year}{2000})
  \bibinfo{pages}{511--528}.
\bibitem[{Bernard et~al.(1993)Bernard, Gaudin, Haldane, and Pasquier}]{BGHP93}
\bibinfo{author}{D.~Bernard}, \bibinfo{author}{M.~Gaudin},
  \bibinfo{author}{F.~D.~M. Haldane}, \bibinfo{author}{V.~Pasquier},
  \bibinfo{journal}{J. Phys. A: Math. Gen.} \bibinfo{volume}{26}
  (\bibinfo{year}{1993}) \bibinfo{pages}{5219--5236}.
\bibitem[{Basu-Mallick et~al.(2010)Basu-Mallick, Bondyopadhaya, and
  Hikami}]{BBH10}
\bibinfo{author}{B.~Basu-Mallick}, \bibinfo{author}{N.~Bondyopadhaya},
  \bibinfo{author}{K.~Hikami}, \bibinfo{journal}{SIGMA} \bibinfo{volume}{6}
  (\bibinfo{year}{2010}) \bibinfo{pages}{091(13)}.
\bibitem[{Haldane(1991{\natexlab{a}})}]{Ha91b}
\bibinfo{author}{F.~D.~M. Haldane}, \bibinfo{journal}{Phys. Rev. Lett.}
  \bibinfo{volume}{67} (\bibinfo{year}{1991}{\natexlab{a}})
  \bibinfo{pages}{937--940}.
\bibitem[{Haldane(1991{\natexlab{b}})}]{Ha91}
\bibinfo{author}{F.~D.~M. Haldane}, \bibinfo{journal}{Phys. Rev. Lett.}
  \bibinfo{volume}{66} (\bibinfo{year}{1991}{\natexlab{b}})
  \bibinfo{pages}{1529--1532}.
\bibitem[{Basu-Mallick et~al.(2008)Basu-Mallick, Bondyopadhaya, and
  Sen}]{BBS08}
\bibinfo{author}{B.~Basu-Mallick}, \bibinfo{author}{N.~Bondyopadhaya},
  \bibinfo{author}{D.~Sen}, \bibinfo{journal}{Nucl. Phys. B}
  \bibinfo{volume}{795} (\bibinfo{year}{2008}) \bibinfo{pages}{596--622}.
\bibitem[{Cirac and Sierra(2010)}]{CS10}
\bibinfo{author}{J.~I. Cirac}, \bibinfo{author}{G.~Sierra},
  \bibinfo{journal}{Phys. Rev. B} \bibinfo{volume}{81} (\bibinfo{year}{2010})
  \bibinfo{pages}{104431(4)}.
\bibitem[{Nielsen et~al.(2011)Nielsen, Cirac, and Sierra}]{NCS11}
\bibinfo{author}{A.~E.~B. Nielsen}, \bibinfo{author}{J.~I. Cirac},
  \bibinfo{author}{G.~Sierra}, \bibinfo{journal}{J. Stat. Mech.-Theory E.}
  (\bibinfo{year}{2011}) \bibinfo{pages}{P11014(39)}.
\bibitem[{Finkel and Gonz{\'a}lez-L{\'o}pez(2005)}]{FG05}
\bibinfo{author}{F.~Finkel}, \bibinfo{author}{A.~Gonz{\'a}lez-L{\'o}pez},
  \bibinfo{journal}{Phys. Rev. B} \bibinfo{volume}{72} (\bibinfo{year}{2005})
  \bibinfo{pages}{174411(6)}.
\bibitem[{Basu-Mallick and Bondyopadhaya(2006)}]{BB06}
\bibinfo{author}{B.~Basu-Mallick}, \bibinfo{author}{N.~Bondyopadhaya},
  \bibinfo{journal}{Nucl. Phys. B} \bibinfo{volume}{757} (\bibinfo{year}{2006})
  \bibinfo{pages}{280--302}.
\bibitem[{Barba et~al.(2008)Barba, Finkel, Gonz\'alez-L\'opez, and
  Rodr{\'\i}guez}]{BFGR08epl}
\bibinfo{author}{J.~C. Barba}, \bibinfo{author}{F.~Finkel},
  \bibinfo{author}{A.~Gonz\'alez-L\'opez}, \bibinfo{author}{M.~A.
  Rodr{\'\i}guez}, \bibinfo{journal}{Europhys. Lett.} \bibinfo{volume}{83}
  (\bibinfo{year}{2008}) \bibinfo{pages}{27005(6)}.
\bibitem[{Barba et~al.(2009)Barba, Finkel, Gonz\'alez-L\'opez, and
  Rodr{\'\i}guez}]{BFGR09}
\bibinfo{author}{J.~C. Barba}, \bibinfo{author}{F.~Finkel},
  \bibinfo{author}{A.~Gonz\'alez-L\'opez}, \bibinfo{author}{M.~A.
  Rodr{\'\i}guez}, \bibinfo{journal}{Nucl. Phys. B} \bibinfo{volume}{806}
  (\bibinfo{year}{2009}) \bibinfo{pages}{684--714}.
\bibitem[{Giuliano et~al.(2010)Giuliano, Sindona, Falcone, Plastina, and
  Amico}]{GSFPA10}
\bibinfo{author}{D.~Giuliano}, \bibinfo{author}{A.~Sindona},
  \bibinfo{author}{G.~Falcone}, \bibinfo{author}{F.~Plastina},
  \bibinfo{author}{L.~Amico}, \bibinfo{journal}{New J. Phys.}
  \bibinfo{volume}{12} (\bibinfo{year}{2010}) \bibinfo{pages}{025022(15)}.
\bibitem[{Kawakami(1992)}]{Ka92}
\bibinfo{author}{N.~Kawakami}, \bibinfo{journal}{Phys. Rev. B}
  \bibinfo{volume}{46} (\bibinfo{year}{1992}) \bibinfo{pages}{1005--1014}.
\bibitem[{Ha and Haldane(1993)}]{HH93}
\bibinfo{author}{Z.~N.~C. Ha}, \bibinfo{author}{F.~D.~M. Haldane},
  \bibinfo{journal}{Phys. Rev. B} \bibinfo{volume}{47} (\bibinfo{year}{1993})
  \bibinfo{pages}{12459--12469}.
\bibitem[{Sutherland(1971{\natexlab{a}})}]{Su71}
\bibinfo{author}{B.~Sutherland}, \bibinfo{journal}{Phys. Rev. A}
  \bibinfo{volume}{4} (\bibinfo{year}{1971}{\natexlab{a}})
  \bibinfo{pages}{2019--2021}.
\bibitem[{Sutherland(1971{\natexlab{b}})}]{Su71b}
\bibinfo{author}{B.~Sutherland}, \bibinfo{journal}{J. Math. Phys.}
  \bibinfo{volume}{12} (\bibinfo{year}{1971}{\natexlab{b}})
  \bibinfo{pages}{246--250}.
\bibitem[{Ha and Haldane(1992)}]{HH92}
\bibinfo{author}{Z.~N.~C. Ha}, \bibinfo{author}{F.~D.~M. Haldane},
  \bibinfo{journal}{Phys. Rev. B} \bibinfo{volume}{46} (\bibinfo{year}{1992})
  \bibinfo{pages}{9359--9368}.
\bibitem[{Calogero(1971)}]{Ca71}
\bibinfo{author}{F.~Calogero}, \bibinfo{journal}{J. Math. Phys.}
  \bibinfo{volume}{12} (\bibinfo{year}{1971}) \bibinfo{pages}{419--436}.
\bibitem[{Minahan and Polychronakos(1993)}]{MP93}
\bibinfo{author}{J.~A. Minahan}, \bibinfo{author}{A.~P. Polychronakos},
  \bibinfo{journal}{Phys. Lett. B} \bibinfo{volume}{302} (\bibinfo{year}{1993})
  \bibinfo{pages}{265--270}.
\bibitem[{Inozemtsev(1996)}]{In96}
\bibinfo{author}{V.~I. Inozemtsev}, \bibinfo{journal}{Phys. Scr.}
  \bibinfo{volume}{53} (\bibinfo{year}{1996}) \bibinfo{pages}{516--520}.
\bibitem[{Frahm and Inozemtsev(1994)}]{FI94}
\bibinfo{author}{H.~Frahm}, \bibinfo{author}{V.~I. Inozemtsev},
  \bibinfo{journal}{J. Phys. A: Math. Gen.} \bibinfo{volume}{27}
  (\bibinfo{year}{1994}) \bibinfo{pages}{L801--L807}.
\bibitem[{Banerjee and Basu-Mallick(2012)}]{BB12}
\bibinfo{author}{P.~Banerjee}, \bibinfo{author}{B.~Basu-Mallick},
  \bibinfo{journal}{J. Math. Phys.} \bibinfo{volume}{53} (\bibinfo{year}{2012})
  \bibinfo{pages}{083301}.
\bibitem[{Miles(1960)}]{Mi60}
\bibinfo{author}{E.~P. Miles, Jr}, \bibinfo{journal}{Am. Math. Mon.}
  \bibinfo{volume}{67} (\bibinfo{year}{1960}) \bibinfo{pages}{745--752}.
\bibitem[{Inozemtsev(1990)}]{In90}
\bibinfo{author}{V.~I. Inozemtsev}, \bibinfo{journal}{J. Stat. Phys.}
  \bibinfo{volume}{59} (\bibinfo{year}{1990}) \bibinfo{pages}{1143--1155}.
\bibitem[{Finkel and Gonz{\'a}lez-L{\'o}pez(2014)}]{FG14JSTAT}
\bibinfo{author}{F.~Finkel}, \bibinfo{author}{A.~Gonz{\'a}lez-L{\'o}pez},
  \bibinfo{journal}{J. Stat. Mech.-Theory E.}  (\bibinfo{year}{2014})
  \bibinfo{pages}{P12014(28)}.
\bibitem[{Nazarov(1991)}]{Na91}
\bibinfo{author}{M.~L. Nazarov}, \bibinfo{journal}{Lett. Math. Phys.}
  \bibinfo{volume}{21} (\bibinfo{year}{1991}) \bibinfo{pages}{123--131}.
\bibitem[{Basu-Mallick et~al.(2007)Basu-Mallick, Bondyopadhaya, Hikami, and
  Sen}]{BBHS07}
\bibinfo{author}{B.~Basu-Mallick}, \bibinfo{author}{N.~Bondyopadhaya},
  \bibinfo{author}{K.~Hikami}, \bibinfo{author}{D.~Sen},
  \bibinfo{journal}{Nucl. Phys. B} \bibinfo{volume}{782} (\bibinfo{year}{2007})
  \bibinfo{pages}{276--295}.
\bibitem[{Polychronakos(1994)}]{Po94}
\bibinfo{author}{A.~P. Polychronakos}, \bibinfo{journal}{Nucl. Phys. B}
  \bibinfo{volume}{419} (\bibinfo{year}{1994}) \bibinfo{pages}{553--566}.
\bibitem[{Basu-Mallick et~al.(1999)Basu-Mallick, Ujino, and Wadati}]{BUW99}
\bibinfo{author}{B.~Basu-Mallick}, \bibinfo{author}{H.~Ujino},
  \bibinfo{author}{M.~Wadati}, \bibinfo{journal}{J. Phys. Soc. Jpn.}
  \bibinfo{volume}{68} (\bibinfo{year}{1999}) \bibinfo{pages}{3219--3226}.
\bibitem[{Barba et~al.(2010)Barba, Finkel, Gonz\'alez-L\'opez, and
  Rodr{\'\i}guez}]{BFGR10}
\bibinfo{author}{J.~C. Barba}, \bibinfo{author}{F.~Finkel},
  \bibinfo{author}{A.~Gonz\'alez-L\'opez}, \bibinfo{author}{M.~A.
  Rodr{\'\i}guez}, \bibinfo{journal}{Nucl. Phys. B} \bibinfo{volume}{839}
  (\bibinfo{year}{2010}) \bibinfo{pages}{499--525}.
\bibitem[{Finkel and Gonz{\'a}lez-L\'opez(2014)}]{FG14}
\bibinfo{author}{F.~Finkel}, \bibinfo{author}{A.~Gonz{\'a}lez-L\'opez},
  \bibinfo{journal}{Ann. Phys.-New York} \bibinfo{volume}{351}
  (\bibinfo{year}{2014}) \bibinfo{pages}{797--827}.
\bibitem[{Spickerman and Joyner(1984)}]{SJ84}
\bibinfo{author}{W.~R. Spickerman}, \bibinfo{author}{R.~N. Joyner},
  \bibinfo{journal}{Fibonacci Quart.}  (\bibinfo{year}{1984})
  \bibinfo{pages}{327--331}.
\bibitem[{Dresden and Du(2014)}]{DD14}
\bibinfo{author}{G.~P.~B. Dresden}, \bibinfo{author}{Z.~Du},
  \bibinfo{journal}{J. Integer Seq.} \bibinfo{volume}{17}
  (\bibinfo{year}{2014}) \bibinfo{pages}{14.4.7(9)}.
\bibitem[{Leinaas and Myrheim(1977)}]{LM77}
\bibinfo{author}{J.~M. Leinaas}, \bibinfo{author}{J.~Myrheim},
  \bibinfo{journal}{Nuovo Cimento} \bibinfo{volume}{37B} (\bibinfo{year}{1977})
  \bibinfo{pages}{1--23}.
\bibitem[{Wilczek(1982{\natexlab{a}})}]{Wi82}
\bibinfo{author}{F.~Wilczek}, \bibinfo{journal}{Phys. Rev. Lett.}
  \bibinfo{volume}{48} (\bibinfo{year}{1982}{\natexlab{a}})
  \bibinfo{pages}{2--4}.
\bibitem[{Wilczek(1982{\natexlab{b}})}]{Wi82b}
\bibinfo{author}{F.~Wilczek}, \bibinfo{journal}{Phys. Rev. Lett.}
  \bibinfo{volume}{49} (\bibinfo{year}{1982}{\natexlab{b}})
  \bibinfo{pages}{957--959}.
\bibitem[{Polychronakos(2006)}]{Po06}
\bibinfo{author}{A.~P. Polychronakos}, \bibinfo{journal}{J. Phys. A: Math.
  Gen.} \bibinfo{volume}{39} (\bibinfo{year}{2006})
  \bibinfo{pages}{12793--12845}.
\bibitem[{Murthy and Shankar(1994)}]{MS94a}
\bibinfo{author}{M.~V.~N. Murthy}, \bibinfo{author}{R.~Shankar},
  \bibinfo{journal}{Phys. Rev. Lett.} \bibinfo{volume}{72}
  (\bibinfo{year}{1994}) \bibinfo{pages}{3629--3633}.
\bibitem[{Beisert(2012)}]{Be12}
\bibinfo{editor}{N.~Beisert} (Ed.), \bibinfo{title}{Special {V}olume: {R}eview
  on {A}dS/{CFT} {I}ntegrability}, volume~\bibinfo{volume}{99} of
  \textit{\bibinfo{series}{Lett. Math. Phys.}}, \bibinfo{publisher}{Springer},
  \bibinfo{year}{2012}.

\end{thebibliography}

\end{document}